%% file: acl_latex.tex
\pdfoutput=1

\documentclass[11pt]{article}

\usepackage[final]{acl}

\usepackage{times}
\usepackage[table]{xcolor}
\usepackage{enumerate}
\usepackage{comment}
\usepackage{booktabs}
\usepackage{enumitem}
\usepackage{latexsym}

\usepackage[T1]{fontenc}

\usepackage[utf8]{inputenc}

\usepackage{microtype}

\usepackage{inconsolata}

\usepackage{graphicx}
\usepackage[edges]{forest}
\usepackage{tikz} 
\usetikzlibrary{shapes,arrows.meta}
\usepackage{float}  

\usepackage{amsmath}
\usepackage{amssymb}
\usepackage{xcolor}
\usepackage{booktabs}
\usepackage{caption}

\title{LLM Agents for Education: Advances and Applications}

\author{
Zhendong Chu\textsuperscript{1}\thanks{Equal contribution.}, 
Shen Wang\textsuperscript{1}\footnotemark[1], 
Jian Xie\textsuperscript{2}\footnotemark[1], 
Tinghui Zhu\textsuperscript{2}\footnotemark[1], 
Yibo Yan\textsuperscript{3}\footnotemark[1], 
Jinheng Ye\textsuperscript{4}\footnotemark[1], \\
\textbf{Aoxiao Zhong\textsuperscript{1},  
Xuming Hu\textsuperscript{3}, 
Jing Liang\textsuperscript{1}, 
Philip S. Yu\textsuperscript{5}, 
Qingsong Wen\textsuperscript{1}} \\
\textsuperscript{1}Squirrel Ai Learning, \textsuperscript{2}Fudan University \\
\textsuperscript{3}Hong Kong University of Science and Technology (Guangzhou) \\
\textsuperscript{4}Tsinghua University, 
\textsuperscript{5}University of Illinois Chicago \\
\texttt{\{zc9uy@virginia.edu, qingsongedu@gmail.com\}}
}

\begin{document}
\maketitle
\begin{abstract}
Large Language Model (LLM) agents are transforming education by automating complex pedagogical tasks and enhancing both teaching and learning processes. In this survey, we present a systematic review of recent advances in applying LLM agents to address key challenges in educational settings, such as feedback comment generation, curriculum design, etc. We analyze the technologies enabling these agents, including representative datasets, benchmarks, and algorithmic frameworks. Additionally, we highlight key challenges in deploying LLM agents in educational settings, including ethical issues, hallucination and overreliance, and integration with existing educational ecosystems. Beyond the core technical focus, we include in Appendix \ref{app:domain} a comprehensive overview of domain-specific educational agents, covering areas such as science learning, language learning, and professional development.
\end{abstract}

\input{tex/introduction}

\input{tex/agent_overview}

\input{tex/Agent4teacher}
\input{tex/Agent4student}

\input{tex/Agent4Challenge}

\input{tex/conclusion_outlook}

\section*{Limitations}
Considering the rapid development of LLM agents for education, it is possible that some of the most recent advancements may not have been captured at the time of writing. Nevertheless, we have made every effort to ensure that all foundational and representative works are included to provide a comprehensive and accurate overview of the field. Specifically, we searched Google Scholar using keywords like ``LLM agents for education'', ``LLM agents for classroom simulation'', ``LLM agents for adaptive learning'', and ``LLM agents for knowledge tracing'' covering publications from 2020 to May 2025. We included studies focusing on LLM agent-based methods but excluded those solely on LLM-based approaches.  

\bibliography{custom}

\clearpage
\appendix

\begin{figure*}
    \centering
    \small
    \resizebox{\linewidth}{!}{
    \input{figure/tax_domain}}
    \caption{Taxonomy of domain-specific educational agents.}
    \label{fig:taxonomy_domain}
    \vspace{-2mm}
\end{figure*}

\section{Domain-Specific Educational Agents}
\label{app:domain}
\input{tex/Agent4Science}
\input{tex/Agent4LanguageLearning}

\input{tex/Agent4ProfessionalDevelopment}
\section{Datasets \& Benchmarks}
\label{sec:benchmark}
In Table \ref{tab:dataset}, we provide a comprehensive summary of publicly available datasets and benchmarks designed to evaluate LLM agents for education across various domains. It categorizes resources based on their primary goal, target users, subject domain, education level, language, modality and dataset size. We hope this collection can support and advance research on LLM agents for education.

Several datasets are designed to evaluate the pedagogical agents, such as ASSIST09 \cite{feng2009addressing} and Junyi \cite{chang2015modeling}, which support knowledge tracing (KT) in K-12 math education, while others like EduAgent \cite{xu2024eduagent} facilitate adaptive learning (AL) by dynamically adjusting content based on student profiles. In addition, error correction and detection (ECD) datasets, such as Virtual Teacher \cite{xu2024ai} and MathCCS \cite{zhang2025correctness}, assess LLM agents' ability to identify and rectify student mistakes in math learning. Other datasets cater to writing, reading, and language learning, including FABRIC \cite{han2023fabric}, EssayJudge \cite{su2025essayjudge}, and EXCGEC \cite{ye2024excgec}, which focus on feedback and generation (FCG) for student essays. MultiSim \cite{ryan-etal-2023-revisiting} and \citet{wang-etal-2024-benchmarking} provide multi-lingual translation and storytelling benchmarks, expanding LLM capabilities beyond English-language education.

Several datasets support domain-specific educational agents across science, law, medicine, and computer science. Beyond their primary goal of evaluating pedagogical ability, these datasets assess LLM agents in domain-specific applications. They provide insights into how LLM agents can be adapted for specialized instruction, evaluating their ability to deliver subject-specific knowledge, facilitate problem-solving, and enhance interactive learning experiences across diverse educational fields.  ScienceAgentBench \cite{chen2024scienceagentbench} and TheoremExplainBench \cite{ku2025theoremexplainagentmultimodalexplanationsllm} assess scientific reasoning and theorem explanation, while ML-Bench \cite{tang2023ml} and MLAgentBench \cite{huang2023mlagentbench} focus on machine learning education. In law education, datasets like LawBench \cite{fei2023lawbench}, LegalBench \cite{guha2023legalbench}, and AgentCourt \cite{chen2024agentcourt} evaluate legal knowledge application, case analysis, and court simulations. Medical education datasets, including MedBench \cite{cai2024medbench} and OmniMedVQA \cite{hu2024omnimedvqa}, test clinical reasoning and medical knowledge retrieval. For computer science, SWE-Bench \cite{yang2025swe} and Programming Feedback \cite{estevez2024evaluation} assess code generation, debugging, and software engineering instruction. These benchmarks help refine LLM agents for specialized tutoring, enhancing AI-driven learning in professional fields.

\input{table/dataset_benchmark}

\end{document}

%% file: tex/introduction.tex
\section{Introduction}
Artificial intelligence techniques are increasingly used in education to enable personalized learning and intelligent tutoring \cite{chen2020artificial, zhai2021review, pedro2019artificial, yan2025position}.  While traditional educational data mining approaches \cite{shafiq2022student, khan2021student, abdelrahman2023knowledge, song2022survey, wang2022neuralcd, gao2021rcd}, such as knowledge tracing and cognitive diagnosis, have made significant progress in reshaping the human-learning paradigm by analyzing student behaviors and assessing knowledge states, they still face major challenges in real-world applications. These challenges include shallow contextual understanding, limited interactive capabilities, and difficulties in generating adaptive, personalized learning materials, etc \cite{zhu2024recommender, laak2024ai, tan2023towards}.


The strong natural language understanding of Large Language Models (LLMs) and the task automation capabilities of LLM agents make them valuable for addressing challenges in education \cite{weng2023prompt, wang2024survey}. First, memory enables LLM agents to retain both long-term knowledge about students' study habits and short-term context from real-time interactions, enhancing contextual understanding and ensuring personalized learning experiences across various educational tasks \cite{zhang2024survey}. Second, tool use allows LLM agents to access external resources, perform complex calculations, and retrieve real-time information, enabling them to automate intricate educational tasks such as grading, knowledge retrieval, and adaptive content generation, thereby overcoming limited interactivity and enhancing engagement \cite{gao2023retrieval}. Third, planning supports structured learning by decomposing complex topics, predicting optimal learning paths, and dynamically adjusting instructional strategies, allowing LLM agents to autonomously guide students through personalized learning experiences \cite{huang2024understanding}. 

In addition to these core architectural capabilities, we identify personalization, explainability, and multi-agent communication as essential for effective educational LLM agents, as they enable nuanced instructional behavior and collaborative reasoning. Personalization allows agents to tailor instruction to individual learners' needs and preferences~\cite{razafinirina2024pedagogical}. Explainability enables them to provide interpretable justifications for feedback and decisions~\cite{abu2024knowledge}. Multi-agent communication facilitates role-based collaboration, such as between a planner and a critic, to improve robustness and task coverage~\cite{wu2023autogen}. By integrating these features, LLM agents enhance understanding and engagement while streamlining educational workflows for greater adaptability and efficiency.

In this survey, we provide a comprehensive review of LLM agents in educational settings, with a focus on their underlying technical foundations and general-purpose pedagogical capabilities. We begin by introducing the core abilities of educational LLM agents, including memory, tool use, planning, personalization, explainability, and multi-agent communication, and discuss their potential to automate and enhance diverse educational tasks. Given the highly applied nature of the education domain, we propose a task-centric taxonomy in Figure~\ref{fig:taxonomy} that organizes recent advances around core educational tasks. We categorize educational LLM agents based on their roles in supporting teachers and students, capturing both instructional and learning-focused functions: (1) Teaching Assistance Agents, which support educators by automating tasks such as \emph{Classroom Simulation (CS)}, \emph{Feedback Comment Generation (FCG)}, and \emph{Curriculum Design (CD)}; and (2) Student Support Agents, which facilitate personalized learning through \emph{Adaptive Learning (AL)}, \emph{Knowledge Tracing (KT)}, and \emph{Error Correction and Detection (ECD)}. The overview of LLM agents for education and illustrative examples of each task are presented in Figure~\ref{fig:overview}. We further highlight critical challenges in deploying these agents, including ethical issues, hallucination and overreliance, and integration into existing educational ecosystems. Finally, we compile essential datasets, benchmarks, and evaluation protocols to support future research on LLM agent-based educational systems. We summarize our contributions as follows:

\begin{itemize}[leftmargin=*]    
    \item \textbf{Novel task-centric taxonomy.} We propose a structured taxonomy that categorizes LLM agents based on core educational tasks, highlighting roles in teaching assistance and student support to unify analysis across applications.

    \item  \textbf{Current challenges and future directions. } We analyze critical challenges that need to be addressed for the effective deployment of LLM agents for education, including issues related to ethical issues, hallucination, and integration into real-world educational ecosystems.

    \item \textbf{Compilation of essential resources. } We compile comprehensive datasets and benchmarks to support future research efforts and facilitate the development of more robust and effective LLM-driven educational solutions.\footnote{Due to the page limit, we present more details in  Appendix \ref{sec:benchmark}}
\end{itemize}

Beyond the main technical focus, Appendix~\ref{app:domain} provides an overview of domain-specific educational agents, including applications in science learning, language learning, and professional development. We outline domain-specific challenges and review recent advances, benchmarks, and datasets. Readers are encouraged to refer to the appendix for further details.

%% file: tex/agent_overview.tex
\begin{figure*}
    \centering
    \small
    \resizebox{\linewidth}{!}{
    \input{figure/taxonomy}} 
    \caption{Taxonomy of representative research on LLM agents for education.}
    \vspace{-2mm}
    \label{fig:taxonomy}
\end{figure*}

\section{LLM Agents for Education}
LLM agents have a set of connected abilities that help them handle complex tasks and provide meaningful support in education. 
These core features work together to let LLM agents do more than just find information---they can interact in ways that are flexible, responsive, and tailored to each learner. 
Their main strengths in education include strong memory, the ability to use tools, planning skills, personalization, clear explanations, and the ability to work with other agents. 
Specifically,

\noindent\textbf{Memory. } Memory in LLM agents includes long-term (foundational knowledge, e.g., commonsense) and short-term (current interaction data) components~\citep{sumerscognitive}.
Active management via summarization and retrieval ensures relevant context is maintained~\citep{chen2023walking,liang2023unleashing,zhong2023memorybank}. This allows agents to track student progress and personalize responses, though sophisticated filtering is crucial to maintain the quality  against noisy interaction data.  

\noindent\textbf{Tool Use. }To overcome limitations like knowledge cutoffs or calculation difficulties, LLM agents use external tools such as search engines, databases, or APIs~\citep{qin2023toolllm,zhuang2023toolqa}. 
This expands their functionality, providing access to current information and specialized capabilities. 
By integrating these tools, LLM agents expand their functional capabilities, ensuring that they can provide accurate and relevant information while supporting diverse educational tasks.

\noindent\textbf{Planning. }Planning allows agents to actively support learning by breaking down complex goals into smaller, manageable steps and adjusting based on how the student is doing~\citep{yao2023tree,valmeekam2023planbench}. 
This means understanding the learning goals, creating step-by-step plans, personalizing the learning path, and making changes based on feedback and progress. 
For longer-term tasks, LLMs can act as ``meta-controllers,'' using methods like ``Pedagogical Steering'' to stay aligned with teaching goals, which thus leads to dynamic, emergent curricula co-created with students~\citep{zhang2025eduplanner}.

\noindent\textbf{Personalization.} Personalization is a hallmark of effective education, and LLM agents excel in this area by adapting to individual learning styles and needs~\citep{chenpersona}. 
They can serve as tutors, teaching assistants, or even simulate peer interactions to enhance the learning experience~\citep{jin2024teachtune,guo2024using}.
Additionally, LLM agents can simulate user behaviors in recommendation systems, which can be applied to personalize educational content~\citep{li2024learning, chu2025uniedu}.
Such a technology of personalization transforms traditional education, making it more accessible and tailored to each student’s unique needs.

\noindent\textbf{Explainability. }Explainability is crucial for building trust and facilitating learning in educational settings~\citep{wu2024usable}. 
LLM agents must provide clear, step-by-step explanations that are understandable to students. 
Clear explanations are particularly important in subjects like STEM, where step-by-step reasoning is essential for student understanding~\citep{nair2024closing,zhang2025sefl}. 
Consequently, the focused effort required to make educational LLM agents more transparent and their decision-making processes more scrutable. 

\noindent\textbf{Multi-Agent Communication. }LLM agents can facilitate richer interactions in collaborative learning environments~\citep{chenpersona, wu2023autogen}. 
Multiple agents could coordinate on group projects or simulate peer learning, fostering critical thinking and diverse perspectives~\citep{jinxin2023cgmi,yue2024mathvc}. 
This approach could support complex simulations of teamwork and real-world professional settings, especially when guided by new teaching frameworks for organization and evaluation~\citep{xu2024eduagent}.


Table~\ref{tab:llm-task-matrix} summarizes the core capabilities required by each educational task supported by LLM agents. For each task, we highlight the primary capabilities identified from the reviewed literature. However, certain studies have also explored additional capabilities beyond these primary ones to enhance specific tasks—for example, \citet{guo2024using} leverage multi-agent communication to improve feedback correctness. We omit these secondary capabilities from the table to more clearly illustrate the primary focus areas associated with each task. 

\begin{figure*}[!t]
    \centering
    \includegraphics[width=1.\textwidth]{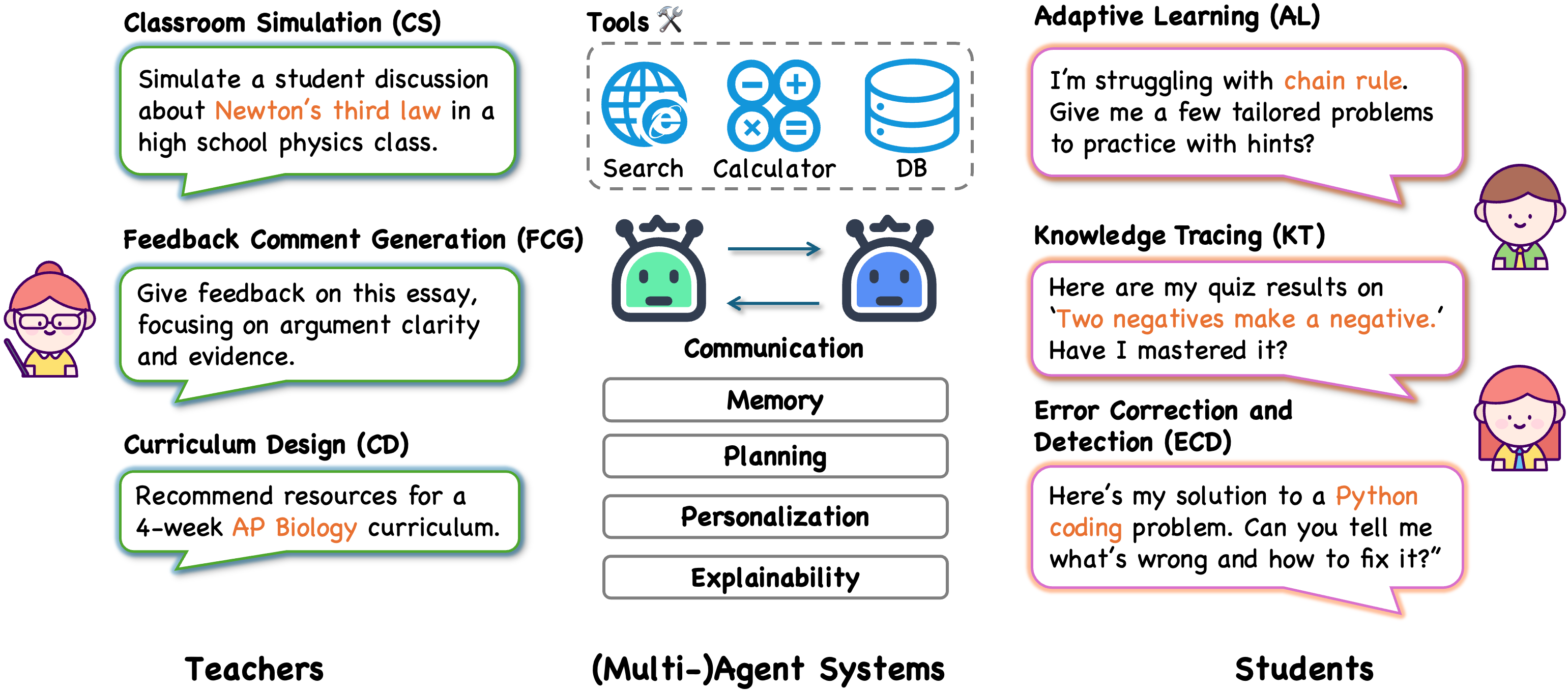}
    \caption{The overview of LLM Agents for education. Teachers and students interact with LLM agents by submitting task-specific prompts. The agents respond using core capabilities such as memory, planning, tool use, and personalization to carry out tasks that support instruction and learning.}
    \label{fig:overview}
\end{figure*}

%% file: figure/taxonomy.tex
\tikzset{
    root/.style =             {align=center, text width=2cm, rounded corners=3pt, line width=0.3mm, fill=gray!10, draw=gray!80, font=\small},
    sa/.style =             {align=center, text width=2cm, rounded corners=3pt, line width=0.3mm, fill=magenta!10, draw=gray!80, font=\small},
    ca/.style =             {align=center, text width=2cm, rounded corners=3pt, line width=0.3mm, fill=green!10, draw=gray!80, font=\small},
    ta/.style =         {align=center, text width=3cm, rounded corners=3pt, line width=0.3mm, fill=blue!10, draw=blue!80, font=\footnotesize},
    ta_work/.style =    {align=center, text width=10cm, rounded corners=3pt, line width=0.3mm, fill=blue!10, draw=blue!0, font=\footnotesize},
    ss/.style =         {align=center, text width=3cm, rounded corners=3pt, line width=0.3mm, fill=red!10, draw=red!80, font=\footnotesize},
    ss_work/.style =    {align=center, text width=10cm, rounded corners=3pt, line width=0.3mm, fill=red!10, draw=red!0, font=\footnotesize},
    sl/.style =           {align=center, text width=3cm, rounded corners=3pt, line width=0.3mm, fill=cyan!10, draw=cyan!80, font=\footnotesize},
    sl_work/.style =      {align=center, text width=10cm, rounded corners=3pt, line width=0.3mm, fill=cyan!10, draw=cyan!0, font=\footnotesize},
    ll/.style =         {align=center, text width=3cm, rounded corners=3pt, line width=0.3mm, fill=orange!10, draw=orange!80, font=\footnotesize},
    ll_work/.style =    {align=center, text width=10cm, rounded corners=3pt, line width=0.3mm, fill=orange!10, draw=orange!0, font=\footnotesize},
    pd/.style =         {align=center, text width=3cm, rounded corners=3pt, line width=0.3mm, fill=purple!10, draw=purple!80, font=\footnotesize},
    pd_work/.style =    {align=center, text width=10cm, rounded corners=3pt, line width=0.3mm, fill=purple!10, draw=purple!0, font=\footnotesize},
}

\begin{forest}
    for tree={
        forked edges,
        grow'=0,
        calign=child edge,
        calign child=(n_children()+1)/2,
    },
    [LLM Agents for Education, root
            [Teaching Assistance \S\ref{sec:teacher}, ta
                    [Classroom Simulation, ta
                        [EduAgent~\citep{xu2024eduagent};
                        TeachTune~\citep{jin2024teachtune};
                        CGMI~\citep{jinxin2023cgmi}
                        Classroom Simulacra~\citep{xu2025classroom};
                        ~\citet{li2025exploring};
                        ~\citet{bhowmik2024evaluation};
                        ~\citet{zheng2025teaching};
                        ~\citet{hu2025exploring},
                        ta_work
                        ]
                    ]
                    [Feedback Comment Generation, ta
                        [
                        PROF~\citep{nair2024closing};
                        SEFL~\citep{zhang2025sefl};
                        FreeText~\citep{matelsky2023large}; 
                        AAAR-1.0~\citep{lou2024aaar};
                        ~\citet{guo2024using};
                        ~\citet{estevez2024evaluation};
                        ~\citet{du2024llms}
                        ,ta_work
                        ]
                    ]
                    [Curriculum Design, ta
                        [\citet{zaiane2002building};
                        SKarREC~\citep{li2024learning};
                        ~\citet{moon2024generative};
                        ~\citet{abu2024knowledge};
                        ~\citet{abu2024supporting}
                        ,ta_work
                        ]
                    ]
                ]
                [Student Support \S\ref{sec:student}, ss
                    [Adaptive Learning, ss
                        [GenAL~\citep{lv2025genal}; EduAgent~\citep{xu2024eduagent};
                        ChatTutor~\citep{chen2024empowering};
                        \citet{park2024empowering};
                        \citet{wang2025llm};
                        \citet{liu-etal-2024-personality};
                        \citet{sonlu2024effects};
                        \citet{liu2024personalized}
                        ,ss_work
                        ]
                    ]
                    [Knowledge Tracing , ss
                        [\citet{yang2024content};
                        \citet{xu2025classroom};
                        \citet{scarlatos2025exploring}
                        ,ss_work
                        ]
                    ]
                    [Error Correction and Detection, ss
                        [\citet{ye2022focus};
                        \citet{li2024rethinking};
                        \citet{xu2024ai};
                        Repairagent~\citep{bouzenia2024repairagent};
                        \citet{wang2025llm};
                        ErrorRadar \citep{yan2024errorradar};
                        CoT Rerailer~\citep{wan2024cot}
                        ,ss_work
                        ]
                    ]
                ]
    ]
\end{forest}

%% file: tex/Agent4teacher.tex

\newcommand{\pending}{\textcolor{gray}{--}}
\begin{table*}[ht]
\centering
\resizebox{\textwidth}{!}{\begin{tabular}{@{}ccccccc@{}}
\toprule
\multicolumn{1}{c}{\textbf{Task}} & \textbf{Memory} & \textbf{Tool Use} & \textbf{Planning} & \textbf{Personalization} & \textbf{Explainability} & \textbf{Multi-Agent Comm.} \\ 
\midrule
CS           & \checkmark & \pending     & \checkmark & \pending         & \pending         & \checkmark \\
FCG   & \pending & \pending     & \pending   & \checkmark       & \checkmark       & \pending    \\
CD              & \checkmark & \checkmark   & \checkmark & \checkmark       & \pending         & \pending    \\ 
\midrule
AL              & \checkmark & \pending  & \checkmark & \checkmark       & \pending         & \pending    \\
KT             & \checkmark & \pending     & \pending   & \checkmark       & \pending         & \pending  \\
ECD & \pending & \checkmark   & \pending   & \checkmark         & \checkmark       & \pending    \\
\bottomrule
\end{tabular}}
\caption{Mapping of educational tasks to essential LLM agent capabilities. $\checkmark$ indicates a primary capability; others may also be used as supporting components in some systems.}
\label{tab:llm-task-matrix}
\end{table*}

\section{Agent for Teaching Assistance}
\label{sec:teacher}
Agents for teaching assistance are designed to support educators in the learning environment. 
These agents leverage LLMs to provide personalized, scalable, and efficient support across various aspects of the educational process. 
Their primary objectives are to enhance teaching quality, enrich student learning experiences, and reduce educators' workload. By incorporating advanced capabilities, LLM agents can effectively support key educational tasks, including classroom simulation (\S\ref{sec:classroom_sim}), feedback comment generation (\S\ref{sec:feedback_comment}), and curriculum design (\S\ref{sec:learning_source}).

\subsection{Classroom Simulation}
\label{sec:classroom_sim}
Classroom simulation refers to the ability of teaching agents to replicate and model various classroom scenarios, such as student-teacher dialogues, collaborative learning activities, and problem-solving tasks. 
These simulations create dynamic and interactive learning environments where educators can experiment with different teaching strategies, assess student reactions, and receive real-time feedback on how various pedagogical approaches may unfold. 
By simulating these classroom dynamics, educators can refine their methods, anticipate student challenges, and enhance overall instructional effectiveness, all without the constraints of a physical classroom setting.

As shown in Table~\ref{tab:llm-task-matrix}, effective classroom simulation relies on memory, planning, and multi-agent communication to accurately model student behavior.
Previous studies~\citep{xu2024eduagent,jin2024teachtune,li2025exploring} demonstrate that LLM-based agents can predict fine-grained student behaviors across diverse personas and past learning patterns, aligning closely with real teachers' expectations. 
To enhance simulation, the CGMI framework~\citep{jinxin2023cgmi} uses a tree-based cognitive architecture with memory, reflection, and planning modules to simulate roles like teacher, student, and supervisor, improving realism. 
Similarly, Classroom Simulacra~\citep{xu2025classroom} incorporates a transferable iterative reflection module for more accurate behavior simulation. 
These systems enable automated interactions that reduce educators' task loads while broadening the exploration of student profiles. 
Simulations can also test educational strategies tailored to different profiles, enhancing teaching quality, as shown in studies by~\citet{bhowmik2024evaluation} and~\citet{zheng2025teaching}. 
Additionally, ~\citet{hu2025exploring} demonstrate how LLMs can refine teaching plans through integrated simulations.
To sum up, classroom simulations can be leveraged to test different educational strategies tailored to diverse student profiles, ultimately enhancing the quality of education. 

\subsection{Feedback Comment Generation}
\label{sec:feedback_comment}

Providing timely, relevant, and constructive feedback is a cornerstone of effective education. 
Teaching assistance agents can generate automated feedback comments on students' assignments, quizzes, and projects. 

For example, \citet{guo2024using} leverage multi-agent communication to provide accurate feedback to students through a two-agent system. Specifically, Agent 1 generates initial feedback based on the students' responses, while Agent 2 evaluates and refines this feedback to prevent over-praise and excessive inferences. 
Furthermore, \citet{nair2024closing} design a training strategy called PROF, which trains an automated LLM-based writing comment generator through reinforcement learning. 
This system adopts an iterative pipeline to simulate various student writing styles and incorporates a more advanced revision model (e.g., GPT-4) to provide the quality of the feedback as rewards.
Similarly, SEFL~\citep{zhang2025sefl} enhances feedback generation by having LLM agents role-play both students and teachers to generate data, which is then used to fine-tune models and improve feedback capabilities. 
These systems have also been deployed in real-world applications, such as FreeText~\citep{matelsky2023large}, which pairs student responses with teacher-provided criteria, enabling the agent to identify strengths and weaknesses and provide targeted feedback for improvement.
Beyond traditional feedback, advanced LLM agents are now capable of handling more complex, expertise-intensive tasks. 
For example, \citet{du2024llms} explores the potential of LLM agents as assistants for natural language processing paper reviewing tasks, while AAAR-1.0~\citep{lou2024aaar} evaluates agents' capabilities in areas such as equation inference, experiment design, paper weakness analysis, and review critique, revealing their potential in conducting research tasks.


However, another line of research highlights that agent-generated feedback still faces challenges in handling complex tasks, such as programming and the review of professional academic papers~\citep{estevez2024evaluation, lou2024aaar}. 
For instance, these agents may struggle with concepts like starvation and deadlocks, leading to inaccurate or incorrect feedback. 
Future work could focus on integrating external tools (e.g., search engines) and enhancing memory mechanisms to better address complex problem-solving scenarios, as well as refining the personalization of feedback across diverse learning contexts.

\subsection{Curriculum Design}
\label{sec:learning_source}
To ensure that students follow personalized learning paths aligned with their knowledge level and domain, it is crucial to develop effective curriculum design strategies. This is a complex task that relies on multiple LLM agent capabilities, including memory, tool use, planning, personalization, and explainability.   

\citet{zaiane2002building} first introduces recommendation systems into e-learning, utilizing web mining techniques to suggest online learning activities or shortcuts on course websites. The integration of LLM-based agents has recently enabled more sophisticated curriculum design through dynamic sequencing and content adaptation. Curricula can be built using either retrieval-based or generation-based methods. Retrieval-based methods involve agents accessing a database or their own memory to suggest existing resources---such as textbooks, research papers, or online content---based on student queries, past behavior, or content similarities~\citep{shahzad2025comprehensive, li2024learning}. 
In contrast, generative methods create new learning content tailored to an individual student's learning style, knowledge gaps, and interests~\citep{moon2024generative}.
Moreover, to enhance the understanding and acceptance of recommended content, these agents need to provide reasons for their recommendations. 
Explaining the rationale behind suggestions fosters trust and enables students to make more informed decisions about the resources they engage with. For example, \citet{abu2024supporting, abu2024knowledge} incorporate knowledge graphs to guide curriculum pathways toward curated and trustworthy sources. This approach not only enhances the interpretability of recommendations but also reduces the risk of generating misinformation, thereby improving the quality of the learning experience.

Looking ahead, future work should focus on integrating agents with adaptive learning systems to dynamically adjust recommendations in response to real-time student performance. In addition, a hybrid approach that combines generative and retrieval-based methods may improve both the accuracy and diversity of curriculum content. Also, incorporating multimodal resources such as interactive media, video, and immersive simulations could further enhance the learning experience. 

%% file: tex/Agent4student.tex
\section{Agent for Student Support}
\label{sec:student}
LLM agent-based student support systems aim to provide real-time personalized assistance without requiring direct teacher involvement. Unlike traditional rule-based systems, these agents offer interactive and adaptive feedback, enabling students to progress at their own pace. Core educational tasks in student support include adaptive learning (\S\ref{sec:adaptive_learning}), knowledge tracing (\S\ref{sec:KT}), and error correction and detection (\S\ref{sec:ECD}).

\subsection{Adaptive Learning}
\label{sec:adaptive_learning}
LLM agents offer the potential to build self-sustaining adaptive learning systems that operate without direct teacher involvement. These systems dynamically tailor instruction based on student performance, enabling personalized learning at scale. Building on the abilities of memory and personalization,  agents are able to maintain and update a structured representation of the learner, commonly referred to as a student profile. This profile informs content selection, pacing, and feedback strategies. 

Several implementations exemplify this adaptive approach. GenAL \cite{lv2025genal} integrates external tools such as automated programs and web searches to construct comprehensive student profiles and inform instructional planning. Based on these profiles, the agent assigns tasks that align with the student’s current knowledge level and updates its memory dynamically.
EduAgent \cite{xu2024eduagent} introduces a structured profiling mechanism comprising four distinct cognitive patterns: gaze behavior linked to physiological memory, motor behavior mapped to motor memory, cognitive state associated with cognitive memory, and post-course assessments contributing to knowledge memory. This structured representation enhances adaptive decision-making by providing a multi-faceted view of student learning states. 
\citet{chen2024empowering} propose a system consisting of interaction, reflection, and reaction, with each component composed of specific LLM tools and memory modules. Furthermore, a meta-agent is introduced to control the information flow through these agents.

Some recent research focuses on modeling students through modular memory or state components that capture \emph{cognitive}, \emph{affective}, and \emph{psychological} dimensions. These representations are often formalized within a state–action framework, where the state space encodes learner traits and the action space governs instructional adaptations. A typical cognitive state representation includes tracking a student’s knowledge proficiency, comprehension levels, and misconceptions \citep{park2024empowering,liu-etal-2024-personality,liu2024personalized}, allowing the agent to tailor explanations, adjust difficulty levels, and reinforce concepts dynamically. Recent studies highlight the importance of affective state modeling, as emotional factors such as motivation, interest, and self-efficacy significantly influence learning outcomes. For instance, \citet{park2024empowering} propose an affective state model that enables agents to adjust feedback tone, provide encouragement, and regulate pacing to maintain engagement. Another crucial dimension of adaptation involves learning preferences and personality traits. These studies integrate personality with memory design, tracking the psychological state of the students. \citet{wang2025llm} integrate learning preferences into state modeling, recognizing that students process information differently depending on instructional format and modality. Adapting content to these preferences enhances retention and learning efficiency. Moreover, \citet{liu-etal-2024-personality} apply the Big Five personality model \citep{roccas2002big} to personalize tutoring strategies, acknowledging that individual differences shape learning experiences \citep{sonlu2024effects}.

Emerging works also explore multi-agent systems for adaptive learning, where specialized agents collaborate to enhance personalization. For example, \citet{wang2025llm} design five agents—Gap Identifier, Learner Profiler, Dynamic Learner Simulator, Learning Path Scheduler, and Content Creator—to deliver goal-oriented, personalized instruction. Similarly, OATutor \cite{10.1145/3544548.3581574} provides an experimental platform for modular adaptive learning, allowing researchers to design scalable, domain-general tutoring agents.

\subsection{Knowledge Tracing}
\label{sec:KT}
Knowledge tracing is essential for monitoring a learner's evolving understanding and predicting future performance. While traditional methods use statistical or deep learning models to estimate mastery, LLM agents offer a more dynamic, personalized approach by leveraging natural language understanding and adaptive instruction. 

Recent advancements have explored multi-agent frameworks for knowledge tracing. For instance, Yang et al. \cite{yang2024content} propose a multi-agent system with three specialized agent roles: administrator, judger, and critic. In this framework, the administrator delegates knowledge tracing tasks to judgers, who collaborate through discussions to assess the student's cognitive state. The critic agent then evaluates the outcome and determines whether the assessment criteria are met, ensuring a structured yet flexible knowledge tracing process. Other agent-based approaches explore alternative strategies for modeling student knowledge. 
\citet{xu2024eduagent} propose simulating students as different personas, allowing agents to adaptively trace knowledge progression based on varied learning profiles. 
Meanwhile, \citet{scarlatos2025exploring} employ dialogue-driven interactions to probe students' conceptual boundaries, using conversational exchanges to refine knowledge estimation dynamically.

\subsection{Error Correction and Detection}
\label{sec:ECD}
Error detection and correction help students refine their understanding through real-time, context-aware feedback. LLM agents can identify mistakes across domains such as writing, programming, and math, and adapt feedback to the learner's proficiency, acting as intelligent reviewers and assistants~\cite{ye2022focus,li2024rethinking}. 

Agent-based systems leverage state representations and adaptive inference mechanisms to track error patterns and misconceptions dynamically \citep{park2024empowering, liu-etal-2024-personality, wang2025llm,bouzenia2024repairagent}. 
Recent advancements extend this capability into the multi-modal domain, incorporating direct analysis of student-generated drafts. 
\citet{xu2024ai} propose a multimodal LLM framework that processes handwritten or digitally drafted student work.
The system first extracts and converts draft content into natural language, enabling the agent to interpret and analyze handwritten responses. 
The agent then provides indirect yet effective instructional feedback, guiding students toward self-correction and deeper comprehension.
Moreover, CoT Rerailer~\cite{wan2024cot} designs a derailment identification process and a rerailment process to conduct error detection when solving math questions. 
\citet{zhang2025correctness} propose the MathCCS benchmark and introduce a sequence error analysis framework that leverages multi-agent collaboration. As the first benchmark for multimodal error detection, ErrorRadar \cite{yan2024errorradar} provides a valuable data foundation for developing multimodal agents in this task. 


%% file: tex/Agent4Challenge.tex
\section{Challenges and Future Directions}
In this section, we discuss key challenges that must be addressed to ensure the effective, reliable, and ethical deployment of LLM agents in educational settings. We focus on three critical areas: (1) privacy, bias, and fairness; (2) hallucination and overreliance; and (3) integration with existing educational ecosystems. For each of these challenges, we outline potential research directions aimed at improving the robustness, trustworthiness, and practical applicability of LLM agents in real-world education environments.

\subsection{Privacy, Bias and Fairness
}

\textbf{Analysis. }LLM agents process vast datasets, often containing sensitive personal information, leading to potential privacy risks. Studies highlight low technological readiness and insufficient privacy measures in educational contexts \cite{yan2024practical}. Emerging research \cite{he2024emerged,gan2024navigating,zhang2024psysafe,hua2024trustagent,huo2025mmunlearner,chen2025safeeraser} underscores new privacy and security concerns, emphasizing the need for stronger data protection mechanisms \cite{huang2023ethics,ismail2025protecting,khan2024ethical}.  Additionally, bias in LLMs remains a pressing concern, as models trained on large datasets can inadvertently reinforce stereotypes and disparities, affecting educational fairness. Recent work calls for bias mitigation strategies to promote equitable learning experiences \cite{adewumi2024fairness,aird2024dynamic,mehrotra2024integrity}. Addressing these biases is essential to ensuring inclusive, unbiased educational outcomes.

\noindent\textbf{Directions. }To overcome the above issues, a number of future directions can be explored: (i) 
\emph{Unlearning for privacy preservation:} leverage advances in machine unlearning \cite{liu2025rethinking} to enable agents to retain useful knowledge while selectively forgetting sensitive user data when required. (ii) \emph{Bias detection and mitigation:} develop automated fairness-checking models that evaluate real-time content generated by LLM agents to detect biased explanations, language, or examples. (iii) \emph{Culturally adaptive LLM Agents for global education:} train multilingual, culturally aware educational agents that dynamically adjust explanations based on regional learning norms, historical perspectives, and diverse curricula.

\subsection{Hallucination and Overreliance}

\textbf{Analysis. }The ``hallucination'' phenomenon, in which LLMs generate plausible but incorrect or nonsensical information, poses a significant challenge to their reliability in educational contexts~\cite{zhang2023siren}. Such inaccuracies can mislead learners by presenting false information with confident and authoritative language, making errors difficult to detect and potentially leading to misconceptions~\cite{da2024mitigation,jho2024leveraging,ho2024mitigating}. For example, AI-generated content may fabricate historical events or scientific facts that students unknowingly accept as true. This risk is further amplified by overreliance, as students and educators may accept AI-generated responses without sufficient critical evaluation. Research has shown that excessive dependence on AI systems can hinder skill acquisition and reduce meaningful engagement with learning materials~\cite{milano2023large, krupp2024challenges, adewumi2023procot}.


\noindent\textbf{Directions. }Some directions can be explored to mitigate hallucinations in LLM agents for education: (i) \emph{Self-correcting AI tutors:} develop LLM agents with self-reflection capabilities \cite{renze2024self}, where models review, verify, and refine their own generated content before presenting it to students. (ii) \emph{Hybrid Human-AI feedback loops for educational content verification:} develop teacher-in-the-loop AI systems where educators can review and correct AI-generated responses, refining agents performance over time. (iii) \emph{Pedagogical-aware educational agents:} design agentic frameworks aligned with human pedagogical expertise to mitigate overreliance on AI-generated content.

\subsection{ Integration  with  Existing Educational Ecosystems}
\textbf{Analysis. }Although LLM agents hold great potential for automating educational practices, it is essential to consider how they can be effectively integrated into existing human-centered educational paradigms. One major challenge is the lack of structured frameworks for integrating LLM agents into educational systems. While models like the FOKE framework \cite{hu2024foke} combine foundation models, knowledge graphs, and prompt engineering to provide interactive and explainable learning services, broader adoption requires scalable models that can be validated in diverse real-world educational settings. Additionally, LLMs have been explored as tools to enhance creativity and collaboration in project-based learning (PBL), supporting students through brainstorming, problem-solving, and project execution. However, studies indicate that their effectiveness is limited by the absence of structured guidance frameworks that help educators and students seamlessly incorporate LLM agents into PBL workflows \cite{zha2024designing}. Another critical challenge is ensuring equitable access to LLM-powered educational tools, particularly in underfunded schools and institutions with limited AI infrastructure. Platforms such as AI-VERDE \cite{mithun2025ai} aim to democratize access by providing a unified \emph{LLM-as-a-platform} service with built-in access control, privacy-preserving mechanisms, and budget management. However, achieving widespread adoption still depends on scalable and cost-effective deployment strategies that can support educational institutions at different resource levels. 

\noindent\textbf{Directions. }Future research should focus on developing standardized frameworks to guide the structured deployment of LLM agents in personalized learning, PBL, and assessment. For example, expanding models like FOKE with adaptive learning strategies, multimodal content, and real-time feedback could enhance instructional effectiveness. Additionally, integrating interactive AI tutors that support student collaboration, project tracking, and contextual guidance would further improve PBL applications. To promote equitable access, develop cost-effective AI tutors through cloud-based and decentralized models would make LLM-powered learning tools more accessible to a wider range of institutions. Finally, to support meaningful integration of LLM agents into educational ecosystems, future work should move beyond task accuracy and explore more practical metrics such as learning gains, user trust, and engagement, which calls for the development of novel education-oriented benchmarks and datasets.

\subsection{Multimodal LLM Agents for Education}
\textbf{Analysis. }Despite rapid progress  \cite{bewersdorff2025taking, kuchemann2025opportunities}, deploying multimodal LLM agents in education faces several unresolved challenges. First, integration of heterogeneous modalities such as speech, images, videos, and VR simulations is technically demanding. Models often overfit to text while underutilizing other modalities \cite{zheng2025mllms}. Second, cultural and contextual alignment is still limited: current agents may misinterpret classroom discourse, gestures, or culturally grounded communication patterns, reducing the accuracy and inclusiveness of feedback. Third, real-time adaptability is critical yet underexplored. For example, VR-based or speech-driven learning environments demand low-latency responses, but existing agents struggle with efficiency when reasoning across modalities \cite{voultsiou2025systematic}. 

\noindent\textbf{Directions. }Some directions can be explored to advance the use of multimodal LLM agents in education: (i) \emph{Robust multimodal fusion:} leverage contrastive pretraining and modality-specific adapters to balance contributions from text, speech, and vision, reducing overfitting to dominant modalities \cite{nguyen2022adaptive}.
(ii) \emph{Context-aware personalization:} incorporate localized knowledge graphs and human-in-the-loop feedback to better align agent reasoning with diverse classroom discourse and cultural norms.
(iii) \emph{Latency-aware optimization:} deploy lightweight distilled models and selective modality routing to ensure real-time responsiveness in VR and speech-driven learning environments.

%% file: tex/conclusion_outlook.tex
\section{Conclusion}
In this survey, we presented a comprehensive review of LLM agents for education, focusing on their technical foundations and their potential to transform personalized learning, intelligent tutoring, and pedagogical automation. We proposed a task-centric taxonomy that categorizes LLM agents into Teaching Assistance and Student Support, highlighting their core capabilities such as memory augmentation, tool use, planning, and personalization. We also examined key research challenges, including ethical issues, hallucination and overreliance, and integration with existing educational ecosystems, which must be addressed to ensure reliable and ethical deployment. To support continued progress in this field, we compiled critical datasets, benchmarks, and evaluation methodologies. As LLM agents continue to evolve, their influence on education will expand. Realizing their full potential, however, will require both technical rigor and thoughtful system design. We hope this survey provides a solid foundation for future research and the development of AI-powered educational systems. 

%% file: figure/tax_domain.tex
\tikzset{
    root/.style =             {align=center, text width=2cm, rounded corners=3pt, line width=0.3mm, fill=gray!10, draw=gray!80, font=\small},
    sa/.style =             {align=center, text width=2cm, rounded corners=3pt, line width=0.3mm, fill=magenta!10, draw=gray!80, font=\small},
    ca/.style =             {align=center, text width=2cm, rounded corners=3pt, line width=0.3mm, fill=green!10, draw=gray!80, font=\small},
    ta/.style =         {align=center, text width=3cm, rounded corners=3pt, line width=0.3mm, fill=blue!10, draw=blue!80, font=\footnotesize},
    ta_work/.style =    {align=center, text width=10cm, rounded corners=3pt, line width=0.3mm, fill=blue!10, draw=blue!0, font=\footnotesize},
    ss/.style =         {align=center, text width=3cm, rounded corners=3pt, line width=0.3mm, fill=red!10, draw=red!80, font=\footnotesize},
    ss_work/.style =    {align=center, text width=10cm, rounded corners=3pt, line width=0.3mm, fill=red!10, draw=red!0, font=\footnotesize},
    sl/.style =           {align=center, text width=3cm, rounded corners=3pt, line width=0.3mm, fill=cyan!10, draw=cyan!80, font=\footnotesize},
    sl_work/.style =      {align=center, text width=10cm, rounded corners=3pt, line width=0.3mm, fill=cyan!10, draw=cyan!0, font=\footnotesize},
    ll/.style =         {align=center, text width=3cm, rounded corners=3pt, line width=0.3mm, fill=orange!10, draw=orange!80, font=\footnotesize},
    ll_work/.style =    {align=center, text width=10cm, rounded corners=3pt, line width=0.3mm, fill=orange!10, draw=orange!0, font=\footnotesize},
    pd/.style =         {align=center, text width=3cm, rounded corners=3pt, line width=0.3mm, fill=purple!10, draw=purple!80, font=\footnotesize},
    pd_work/.style =    {align=center, text width=10cm, rounded corners=3pt, line width=0.3mm, fill=purple!10, draw=purple!0, font=\footnotesize}
}

\begin{forest}
    for tree={
        forked edges,
        grow'=0,
        calign=child edge,
        calign child=(n_children()+1)/2,
    },
    [Domain-Specific Educational Agents, root
        [Science Learning \S\ref{sec:science}, sl
            [Mathematics, sl
                [MathAgent~\citep{yan2025mathagent};
                TORA~\citep{gou2023tora};
                MathChat~\citep{wu2023mathchat};
                MACM~\citep{lei2025macm};
                \citet{xiong2024building},
                sl_work
                ]
            ]
            [Physics, sl
                [NEWTON~\citep{wang2023newton};
                Physics Reasoner~\citep{pang2024physics};
                SGA~\citep{ma2024llm},
                sl_work
                ]
            ]
            [Chemistry, sl
                [ChemCrow~\citep{m2024augmenting};
                ChemAgent~\citep{yu2024tooling};
                ChemReasoner~\citep{sprueill2024chemreasoner};
                Curie~\citep{kon2025curie},
                sl_work
                ]
            ]
            [Biology, sl
                [ProtChat~\citep{huang2024protchat};
                ProtAgents~\citep{ghafarollahi2024protagents};
                TourSynbio-Agent~\citep{shen2024toursynbio},
                sl_work
                ]
            ]
            [General Scientific Discovery, sl
                [PaperQA \cite{lala2023paperqa};
                PaperQA2 \cite{skarlinski2024language};
                OpenScholar \cite{asai2024openscholar};
                SciAgents \cite{ghafarollahi2024sciagents};
                SciAgent \cite{ma2024sciagent};
                SciToolAgent \cite{chen2025scitoolagent};
                TheoremExplainAgent \cite{ku2025theoremexplainagentmultimodalexplanationsllm};
                LLM-SR \cite{shojaee2024llm};
                AI Scientist \cite{lu2024ai};
                ResearchAgent \cite{baek2024researchagent};
                Agent Laboratory \cite{schmidgall2025agent};
                DiscoveryWorld~\citep{jansen2024discoveryworld},
                sl_work
                ]
            ]
        ]
        [Language Learning \S\ref{sec:language}, ll
            [Reading, ll
                [ExpertEase \cite{mo-hu-2024-expertease};
                AgentSimp \cite{fang-etal-2025-collaborative};
                \citet{chen2024llms},
                ll_work
                ]
            ]
            [Writing, ll
                [Weaver \cite{wang2024weaver};
                CAELF \cite{hong2024my};
                EvaAI \cite{lagakis2024evaai};
                Debate-to-Write \cite{hu-etal-2025-debate},
                ll_work
                ]
            ]
            [Translation, ll
                [Agent-SimT \cite{guo2024agent};
                TransAgents \cite{wu2024perhaps},
                ll_work
                ]
            ]
            [Storytelling, ll
                [STARie \cite{li2023designing};
                StoryAgent \cite{sohn2024words};
                LLaMS \cite{zhang2025let},
                ll_work
                ]
            ]
            [Speaking, ll
                [ELLMA-T \cite{pan2024ellma};
                Spoken-WOZ \cite{si2023spokenwoz};
                FurChat \cite{cherakara-etal-2023-furchat},
                ll_work
                ]
            ]
        ]
        [Professional \\Development \S\ref{sec:professional}, pd
            [Medical Education, pd
                [Talk2Care \cite{yang2024talk2care};
                openCHA \cite{abbasian2023conversational};
                MEDCO \citep{wei2024medco};
                \citet{abd2023large};
                \citet{steenstra2024engaging};
                Agent Hospital \cite{li2024agent},
                pd_work
                ]
            ]
            [Computer Science Education, pd
                [CodeAgent \cite{zhang2024codeagent};
                ToolCoder \cite{ding2025toolcoder};
                SWE-agent \cite{yang2025swe};
                AgentCoder \cite{huang2023agentcoder};
                MapCoder~\cite{islam2024mapcoder};
                Iris \cite{bassner2024iris};
                \citet{chen2024learning};
                HypoCompass \cite{ma2024teach};
                AlgoBo \cite{jin2024teach};
                Magentic-One \cite{fourney2024magentic};
                AutoKaggle \cite{li2024autokaggle};
                AgentK v1.0 \cite{grosnit2024large};
                SWE-Search \cite{antoniades2024swe};
                AgentLess \cite{xia2024agentless};
                TRAVER~\citep{wang2025training},
                pd_work
                ]
            ]
            [Law Education, pd
                [Lawbench \cite{fei2023lawbench};
                LegalBench \cite{guha2023legalbench};
                LegalAgentBench \citep{li2024legalagentbench};
                DeliLaw \citep{xie2024delilaw};
                Agentcourt \citep{chen2024agentcourt};
                Agentscourt \citep{he2024agentscourt};
                AgentsBench \citep{jiang2024agents};
                LawLuo \citep{sun2024lawluo},
                pd_work
                ]
            ]
        ]
    ]
\end{forest}

%% file: tex/Agent4Science.tex
Recent research on LLM agents in education has also shown growing interest in domain-specific applications. We explore their use in \emph{science learning}, \emph{language learning}, and \emph{professional development}, focusing on their algorithmic frameworks, agentic designs, and relevant datasets and benchmarks. In Figure \ref{fig:taxonomy_domain}, we present a systematic taxonomy of domain-specific educational agents.

\subsection{Agent for Science Learning}
\label{sec:science}
An agent for science learning is an intelligent system powered by LLMs, designed to assist students in acquiring and applying scientific knowledge through personalized, interactive experiences \citep{yan2025position,raihan2025large,ng2024empowering,brown2025coordinate}. The significance of these agents in education lies in their ability to offer tailored feedback, enhance conceptual understanding, and promote active engagement with complex scientific ideas. In the following sections, we explore the impact of LLM agents in four key scientific disciplines: \emph{mathematics} (\S\ref{sec:math}), \emph{physics} (\S\ref{sec:physics}), \emph{chemistry} (\S\ref{sec:chem}), and \emph{biology} (\S\ref{sec:bio}), as well as their broader contributions to \emph{general scientific discovery} (\S\ref{sec:general_science}).

\subsubsection{Mathematics}
\label{sec:math}
In mathematics, LLM agents provide substantial support by helping students navigate complex problems and reinforcing their understanding of abstract concepts \citep{yan2024survey,xiong2024building,swan2023math,yan2024errorradar,wu2025empirical,mitra2024orca}. For instance, \citet{gou2023tora} introduce TORA (Tool-integrated Reasoning Agents), a framework that integrates natural language reasoning and program-based tool use to handle mathematical reasoning. MathAgent \citep{yan2025mathagent} similarly proposes Mixture-of-Math-Agent framework to address multimodal error detection in real-world K-12 scenarios, and flexibly transform the visual information of different types of questions into forms that are more easily understood by LLMs (\textit{e.g.,} converting plane geometry images into formalized expression). Additionally, MathChat \citep{wu2023mathchat} serves as a conversational mathematical problem-solving agent, which consists of a chat-based LLM agent and a tool-based user agent. Furthermore, \citet{xiong2024building} propose to use reinforcement learning from human feedback (RLHF) to further improve tool-integrated agents for mathematical problem-solving, and formulate this method as a Markov decision process, distinguishing it from the typical contextual bandit approach used in RLHF. Besides, MACM \citep{lei2025macm} discuss the limitations of LLMs in handling complex mathematical logical deduction, thus introducing a multi-agent system, which comprises three interactive agents: Thinker, Judge, and Executor.

\subsubsection{Physics}
\label{sec:physics}
In the field of physics, LLM agents help students make sense of challenging concepts and offer interactive tools to simulate physical phenomena \citep{pang2024physics,mower2025khwarizmi,barman2025large,fengphyspde,jiang2024beyond,yan2024georeasoner}.
\citet{wang2023newton} introduce NEWTON, the first pipeline and benchmark to explore the physical reasoning abilities of LLMs.
Furthermore, \citet{kortemeyer2023could} describe a case study exploring if an LLM agent can pass an introductory calculus-based physics course.
In addition, Physics Reasoner \citep{pang2024physics}, a novel knowledge-augmented framework for physics problem-solving, leverages a comprehensive formula set and detailed checklists to ensure accuracy and completeness. It can serve as an agent consisting of three stages - problem analysis, formula retrieval, and guided reasoning.
Besides, \citet{ma2024llm} describe the Scientific Generative Agent (SGA), a bilevel optimization framework designed for physical scientific discovery, and highlight the use of LLMs for generating and revising scientific hypotheses and implementing an exploit-and-explore strategy.

\subsubsection{Chemistry}
\label{sec:chem}
Chemistry education also benefits greatly from LLM agents, which can explain molecular structures, chemical reactions, and experimental processes in an engaging and interactive way \citep{ramos2025review,m2024augmenting,yu2024tooling,guo2023can,tsai2023exploring}. 
For example, ChemCrow \citep{m2024augmenting} is the first LLM chemistry agent capable of autonomous planning and execution of chemical syntheses, including an insect repellent and three organocatalysts.
\citet{yu2024tooling} further present ChemAgent, an enhanced chemistry agent improved over ChemCrow, with a focus on two essential cognitive abilities of chemistry problem-solving: reasoning and grounding.
Besides, Curie \citep{kon2025curie} is an agent framework aimed at incorporating rigor into the experimentation process via three core elements: an intra-agent rigor module to boost reliability, an inter-agent rigor module to ensure systematic control, and an experiment knowledge module to improve interoperability.
Recent studies have explored the capabilities of LLMs in complex chemical discovery \citep{yang2024moose,ruan2024accelerated,sprueill2024chemreasoner,moret2023leveraging,jablonka2024leveraging}, and their potential can be advanced by leveraging the interactivity of agent-based tool use and the flexibility of planning strategies \citep{song2024multi,ramos2025review}.

\subsubsection{Biology}
\label{sec:bio}
In biology, LLM agents enhance learning by offering detailed explanations of biological processes and providing interactive experiences to explore living systems \citep{yan2025position,bhattacharya2024large,sripathi2024machine,zhao2025biomaze,gao2024empowering}. 
For example, ProtChat \citep{huang2024protchat} is a multi-agent tool leveraging GPT-4 and Protein Language Models for seamless protein analysis automation, thus evolutionizing the complexities of protein sequence interpretation. 
ProtAgents \citep{ghafarollahi2024protagents} is introduced as a multi-agent modeling framework that combines state-of-the-art LLMs with diverse tools to tackle protein design and analysis. It consists of a team of agents: User, Planner, Assistant, Critic, and Group Chat Manager.
Besides, \citet{shen2024toursynbio} present TourSynbio-Agent, an innovative agent framework that leverages TourSynbio-7B's protein understanding ability to perform various protein engineering tasks, such as mutation analysis, inverse folding, and visualization.

\subsubsection{General Scientific Discovery}
\label{sec:general_science}
LLM agents support general scientific discovery by assisting students in data interpretation, hypothesis testing, and creative problem-solving \citep{yan2025position,chen2024scienceagentbench,ghafarollahi2024sciagents,chen2025scitoolagent,schmidgall2025agent}. 
These LLM agents, such as PaperQA \citep{lala2023paperqa}, PaperQA2 \citep{skarlinski2024language}, OpenScholar \citep{asai2024openscholar}, SciAgents \citep{ghafarollahi2024sciagents}, TheoremExplainAgent \citep{ku2025theoremexplainagentmultimodalexplanationsllm}, and LLM-SR \citep{shojaee2024llm}, can analyze complex scientific datasets, helping students uncover patterns and trends that may not be immediately apparent. 
In addition, \citet{narayanan2024aviary} present Aviary, an extensible gymnasium for language agents for three challenging scientific tasks: manipulating DNA constructs for molecular cloning, answering research questions by accessing scientific literature, and engineering protein stability.
Furthermore, both SciAgent \citep{ma2024sciagent} and SciToolAgent \citep{chen2025scitoolagent} extend to a tool-augmented scientific reasoning setting with the help of domain-specific tools.
Besides, Agent Laboratory \citep{schmidgall2025agent} emerges as an agent framework that automates the research process of three phases (Literature Review, Experimentation, and Report Writing) via various LLM agents (PhD, Postdoc, ML Engineer, \textit{etc.}).

%% file: tex/Agent4LanguageLearning.tex
\subsection{Agent for Language Learning}
\label{sec:language}
The integration of LLM agents into language learning is revolutionizing how core competencies—reading, writing, listening, and speaking—are taught and practiced~\cite{ye2025position}. These skills form the foundation of effective communication and language acquisition, and recent advancements in LLM-based agents have significantly enhanced how learners interact with and acquire these skills. Below, we introduce recent advancements in each subdomain, highlighting the role of LLM agents in enhancing pedagogical outcomes in language acquisition for students and second language (L2) speakers through engaging and adaptive approaches~\cite{huang2023frustratingly,ye2023system,ye2023mixedit}.

\subsubsection{Reading}
Reading comprehension is a vital component of language learning, and LLM agents are playing an increasingly important role in enhancing students' reading abilities. For instance, ExpertEase~\cite{mo-hu-2024-expertease} employs a multi-agent framework to adapt documents for grade-specific audiences, simulating expert-teacher-student collaboration to enhance comprehension. AgentSimp \cite{fang-etal-2025-collaborative} tackles document-level simplification by leveraging multiple agents with distinct roles to ensure coherence and accessibility. Additionally, various LLMs~\cite{adetayo2024microsoft} have been used as academic reading companions, demonstrating improved engagement and understanding of complex qualitative texts in educational settings~\cite{chen2024llms}.

\subsubsection{Writing}
The development of writing skills has benefited significantly from NLP tasks like explainable grammatical error correction (EXGEC)~\citep{ye2024excgec,ye2025corrections,zou2025revisiting} and automatic essay scoring (AES) systems~\citep{su2025essayjudge}. Weaver~\citep{wang2024weaver}, a family of LLMs fine-tuned for writing tasks~\cite{ye2023cleme,ye2024cleme2}, outperforms generalist LLMs like GPT-4 in generating human-like narratives. Moreover, Weaver natively supports retrieval-augmented generation (RAG) and function calling, serving as a qualified foundational model for LLM agents. For interactive feedback on student essays, CAELF~\cite{hong2024my} introduces a multi-agent framework that enables interactive essay feedback. By combining Teaching-Assistant agents' evaluations with teacher-agent arbitration, students can contest grades and engage with the feedback, addressing the ``black box'' limitations of traditional automated scoring.  Inspired by the process of human debate, Debate-to-Write~\cite{hu-etal-2025-debate} construct a persona-based multi-agent framework that can enable agents to collaboratively debate, discuss ideas, and form a comprehensive plan for argument writing.

\subsubsection{Translation}
LLM agents demonstrate remarkable advancements in both simultaneous~\cite{koshkin2024transllama,guo2024sillm} and literary translation~\cite{cheng2024towards,vskobo2023navigating}. Translation tasks benefit from LLM agents through their ability to integrate specialized tools and orchestrate multi-agent collaboration. Agent-SiMT~\cite{guo2024agent} combines the decision-making capabilities of a Simultaneous Machine Translation (SiMT) policy agent with the generative power of a translation agent, achieving state-of-the-art performance in simultaneous translation by dynamically balancing reading and generation actions. For literary translation, TransAgents~\cite{wu2024perhaps} employs a multi-agent framework to replicate the complex workflows of human translation teams, addressing cultural nuances and stylistic challenges through collaborative reasoning. This approach not only improves translation quality but also extends LLM applications to linguistically and culturally rich domains. These contributions underscore the importance of tool use and agent collaboration in advancing translation education~\cite{zheng2024fine}.

\subsubsection{Storytelling}
Storytelling applications leverage LLM agents to create immersive and interactive learning experiences~\cite{simon2022tattletale}. STARie~\cite{li2023designing}, a peer-like embodied conversational agent, integrates multimodal tools such as speech synthesis and facial animation to scaffold children’s storytelling, fostering narrative creativity and oral communication skills~\cite{beredo2021beyond,cassell2022socially}. StoryAgent~\cite{sohn2024words} combines top-down story drafting with bottom-up asset generation to transform simple prompts into coherent, multi-modal digital narratives. By automating complex storytelling workflows~\cite{liem2023workflow}, it democratizes content creation and enhances engagement in language learning. LLaMS~\cite{zhang2025let}, a multi-modal agent framework, is designed to generate multi-modal human-level stories characterized by expressiveness and consistency, incorporating the Story-Adapter module for long image sequence illustration. These systems demonstrate the potential of LLM agents to support both cognitive and creative aspects of language education for children.

\subsubsection{Speaking}
LLM agents are revolutionizing spoken language education by integrating reasoning and multi-agent collaboration to build adaptive dialogue systems~\cite{liu2024understanding,balan2024use}. ELLMA-T~\cite{pan2024ellma} employs contextual reasoning and role-playing in social VR environments to provide personalized feedback and language assessments, enabling learners to practice speaking in realistic scenarios~\cite{lim2024artificial,li2025embodied}. SpokenWOZ~\cite{si2023spokenwoz} introduces a large-scale benchmark for task-oriented spoken dialogue, highlighting the importance of reasoning and multi-turn interaction in addressing real-world conversational challenges. FurChat~\cite{cherakara-etal-2023-furchat}, an embodied conversational agent, combines verbal and non-verbal communication cues to simulate natural interactions, making it a valuable tool for improving speaking skills through immersive and realistic practice. By employing multimodal signals such as speech and gestures, SpeechAgents~\cite{zhang2024speechagents} enhances the authenticity of dialogue simulations, capturing consistent content, natural rhythm, and rich emotional expression. Through Multi-Agent Tuning~\cite{liang2024cmat}, it optimizes LLM capabilities for large-scale simulations involving up to 25 agents, enabling applications like drama creation and audio novel generation.







%% file: tex/Agent4ProfessionalDevelopment.tex
\subsection{Agent for Professional Development}
\label{sec:professional}
Agents for professional development harness the capabilities of LLMs to offer scalable, adaptive, and context-aware learning experiences tailored to domain-specific needs. This section summarizes how recent studies develop agents to revolutionize professional training in fields including \textit{medical} 
 (\S\ref{subsubsec:medical}), \textit{computer science} (\S\ref{subsubsec:cs}), and \textit{law education} (\S\ref{subsubsec:law}).

\subsubsection{Medical Education}\label{subsubsec:medical}
The deployment of LLM agents in healthcare has created new opportunities for personalized, interactive, and scalable systems, with several health agents introduced~\cite{shusterman2025active} such as Talk2Care~\cite{yang2024talk2care} and openCHA~\cite{abbasian2023conversational}. Additionally, \citet{abd2023large} highlight the educational potentials of LLMs in crafting personalized curricula, adaptive learning plans, and dynamic assessment tools for medical education, while concurrently addressing challenges including algorithmic bias, misinformation, and privacy issues. MEDCO~\cite{wei2024medco}, a multi-agent system, has the capacity to replicate real-world medical training environments through agent collaboration with virtual patients, expert physicians, and radiologists, enhancing interdisciplinary learning and peer interaction. Furthermore, \citet{abbasian2023conversational} introduce openCHA as a personalized LLM-powered framework that integrates external resources and orchestrates multi-step problem-solving for complex healthcare queries~\cite{ye2024productagent}, emphasizing tool use and action planning. Beyond traditional education, \citet{steenstra2024engaging} explore LLMs in creating fantasy narrative games for adolescent health education, demonstrating the agents’ ability to generate engaging, doctor-validated content that enhances knowledge retention through gamification. \citet{li2024agent} present Agent Hospital, a simulation environment where LLM-driven agents evolve through autonomous interactions, demonstrating significant improvements in medical reasoning and performance on benchmarks like MedQA~\cite{jin2021disease} after treating thousands of simulated patients. Collectively, these investigations highlight the versatility of LLM agents within medical education, demonstrating their abilities in reasoning, collaboration, tool integration, and adaptive learning to effectively address a broad spectrum of educational and clinical challenges~\cite{karabacak2023embracing,tian2024opportunities,ullah2024challenges}.

\subsubsection{Computer Science Education}\label{subsubsec:cs}
An agent for computer science (CS) education greatly enhances learning by providing personalized guidance on coding, debugging, and understanding CS principles \cite{ma2024teach,lee2024teachers,kosar2024computer,liu2024teaching}. 
For example, CodeAgent \cite{zhang2024codeagent} serves as an LLM agent framework for repo-level code generation, incorporating external tools such as WebSearch and DocSearch. 
Recent studies have demonstrated the potential of agent-based code generation systems such as ToolCoder \cite{ding2025toolcoder}, SWE-agent \cite{yang2025swe}, AgentCoder \cite{huang2023agentcoder}, and MapCoder \cite{islam2024mapcoder}, which can significantly enhance students' coding efficiency \cite{jin2024teach,wang2025training,frankford2024survey}.
Furthermore, \citet{bassner2024iris} introduce Iris, an LLM-driven virtual tutor designed to offer personalized, context-aware assistance to CS students within the interactive learning platform Artemis.
Besides, \citet{chen2024learning} propose Learning-by-Teaching (LBT) as an effective pedagogical strategy for CS education, and leverage the advantages of LLM agents (\textit{e.g.,} contextual conversation \& learning from demonstrations).

\subsubsection{Law Education}\label{subsubsec:law}
LLM agents leverage pre-trained legal knowledge, interactive capabilities, and reasoning skills to support law education through judicial interpretation, moot court simulation, and case analysis \cite{chen2024agentcourt, nelson2024other, lai2024large, yuan2024can}. However, evaluations from LawBench \cite{fei2023lawbench} and LegalBench \cite{guha2023legalbench} reveal that LLMs struggle with legal knowledge application and judicial aid. LegalAgentBench \citep{li2024legalagentbench} further highlights their limitations in multi-hop reasoning and defense statement writing, showing that LLM agents require significant improvements to effectively assist in complex legal tasks. Despite these challenges, LLM agents are emerging as valuable tools for moot court simulations, a crucial component of legal reasoning and advocacy training. DeliLaw \cite{xie2024delilaw} enhances law education by integrating legal and case retrieval modules, enabling students to practice legal research, case analysis, statutory interpretation, and mock consultations. LawLuo \cite{sun2024lawluo} applies a multi-agent framework with retrieval-augmented generation to simulate multi-turn legal consultations, improving personalization and ambiguity handling. Similarly, AgentCourt \cite{chen2024agentcourt} and AgentsCourt \cite{he2024agentscourt} simulate courtroom interactions and judicial decision-making, providing a realistic training ground for law students. AgentsBench \cite{jiang2024agents} extends this by offering multi-agent legal reasoning and case analysis, further advancing AI-driven legal education. 

%% file: table/dataset_benchmark.tex
\begin{table*}
\centering
\caption{Summary of existing datasets and benchmarks of LLM agents for education.}
\label{tab:dataset}
\resizebox{\textwidth}{!}{
\begin{tabular}{ccccccccc}
\toprule
    Dataset\&Benchmark & Goal & User & Domain & Level & Language & Modality & Amount & Source \\
\midrule
ASSIST09 & KT & Student & Math & K12 &  EN & text & 227k & \cite{feng2009addressing} \\
Junyi & KT & Student & Math & K12 & ZH & text & 2.5M & \cite{chang2015modeling} \\
EduAgent & AL \& CS & Student & - & Graduate & EN & text \& image & 1,015 & \cite{xu2024eduagent} \\
MathDial & AL \& KT & Student & Math & K12 & EN & text & 45 & \cite{macina2023mathdial} \\
MultiArith & AL & Student & Math & K12 & EN & text & 180 & \cite{xu2024eduagent} \\
CoMTA & KT & Student & Math & K12 & EN & text & 153 & \cite{scarlatos2025exploring} \\
MaCKT & KT & Student & Math & K12 & EN & text & 452 & \cite{yang2024content} \\
Virtual Teacher & ECD & Student & Math & K12 & ZH & text \& image & 420 & \cite{xu2024ai} \\
MathCCS & ECD & Student & Math & K12 & ZH & text \& image & 420 & \cite{zhang2025correctness} \\
MathTutorBench & ECD \& FCG & Student & Math & K12 & EN & text & 7 tasks & \cite{macina2025mathtutorbench}
\\ \hline

MultiSim & - & Student & Reading & - & Multi-lingual & text & 1.7M & \cite{ryan-etal-2023-revisiting} \\
FABRIC & FCG & Student & Writing & - & EN & text & 1,782 & \cite{han2023fabric} \\
EssayJudge & FCG & Student & Writing & - & EN & text \& image & 1,054 & \cite{su2025essayjudge} \\ 
PROF & FCG & Teacher & Writing & - & EN & text & 363 & \cite{nair2024closing} \\
EXCGEC & FCG & Student & Writing & - & ZH & text & 8,216 & \cite{ye2024excgec} \\
\citet{wang-etal-2024-benchmarking} & - & All & Translation & - & Multi-lingual & text & 70K & \cite{wang-etal-2024-benchmarking} \\
NewEpisode & - & Student & Storytelling & - & EN & text \& image & 24.5K & \cite{wang2024evolving} \\
SD-Eval & ECD & Student & Speaking & - & EN & speech & 7,303 & \cite{ao2024sdeval} \\

\hline

Programming Feedback & FCG & Teacher & Computer Science & - & Code & text & 52 & \cite{estevez2024evaluation}
\\ 
Review Critique & FCG & All &Computer Science& - & EN & text & 440 & \cite{du2024llms}
\\
AAAR-1.0 & FCG & All & Computer Science & - & EN & text \& image & 1,000 & \cite{lou2024aaar}

\\ 
\hline 
ScienceAgentBench & - & All & General Science& - & EN & text & 102 & \cite{chen2024scienceagentbench} \\
TheoremExplainBench & - & All & General Science& - & EN & text & 240 & \cite{ku2025theoremexplainagentmultimodalexplanationsllm} \\
MLGym-Bench & - & All & General Science& - & EN & text & 13 tasks & \cite{nathani2025mlgym} \\
ML-Bench & - & All & Machine Learning& - & EN & text \& image &9,641 & \cite{tang2023ml} \\
MLAgentBench & - & All & Machine Learning& - & EN & text & 13 tasks & \cite{huang2023mlagentbench} \\
SciCode & - & All & General Science& - & EN & text & 338 & \cite{tian2024scicode} \\
BLADE & - & All & General Science& - & EN & text & 12 tasks & \cite{gu2024blade} \\
DiscoveryBench & - & All & General Science& - & EN & text &1,167 & \cite{majumder2024discoverybench} \\
SUPER & - & All & General Science& - & EN & text &796 & \cite{bogin2024super} \\
E-EVAL & - & All & Math \& Language \& General Science & K12 & ZH & text & 4,351 & \cite{hou2024eval} \\

\hline

MedBench & - & All & Medical & - & ZH  & text & 40,041& \cite{cai2024medbench} \\
CMB & - & All & Medical & - & ZH  & text & 280,839 & \cite{wang-etal-2024-cmb} \\
OmniMedVQA & - & All & Medical & - & ZH & text \& image & 127,995 & \cite{hu2024omnimedvqa} \\
MedEval & - & All & Medical & - & EN &  text \& image & 22,779 & \cite{he-etal-2023-medeval} \\

\hline
OSWorld & - & All & Computer Science& - & EN & text \& image & 369 & \cite{xie2024osworld} \\
Spider2-V & - & All & Computer Science& - & EN & text \& image & 812 & \cite{cao2024spider2} \\
VisualWebArena & - & All & Computer Science& - & EN & text \& image & 910 & \cite{koh2024visualwebarena} \\
WebArena & - & All & Computer Science& - & EN & text \& image & 812 & \cite{zhou2023webarena} \\
SWE-Bench & - & All & Computer Science& - & EN & text & 2,294 & \cite{yang2025swe} \\
SWE-Bench M & - & All & Computer Science& - & EN & text \& image & 617 & \cite{yang2024swe} \\
Magentic-One & - & All & Computer Science& - & EN & text \& image & 617 & \cite{fourney2024magentic} \\
\hline
Lawbench & - & All & Law & - & ZH & text & 20 tasks & \cite{fei2023lawbench} \\
LegalBench & - & All & Law & - & EN & text & 162 tasks & \cite{guha2023legalbench} \\
LegalAgentBench & - & All & Law & - & ZH & text & 300 tasks & \cite{li2024legalagentbench} \\
Agentcourt & - & All & Law & - & ZH & text & 550 &  \cite{chen2024agentcourt} \\
SimuCourt & - & All & Law & - & ZH & text & 420 &  \cite{he2024agentscourt} \\

\bottomrule
\end{tabular}}
\end{table*}

%% file: acl_latex.bbl
\begin{thebibliography}{265}
\providecommand{\natexlab}[1]{#1}

\bibitem[{Abbasian et~al.(2023)Abbasian, Azimi, Rahmani, and Jain}]{abbasian2023conversational}
Mahyar Abbasian, Iman Azimi, Amir~M Rahmani, and Ramesh Jain. 2023.
\newblock Conversational health agents: A personalized llm-powered agent framework.
\newblock \emph{arXiv preprint arXiv:2310.02374}.

\bibitem[{Abd-Alrazaq et~al.(2023)Abd-Alrazaq, AlSaad, Alhuwail, Ahmed, Healy, Latifi, Aziz, Damseh, Alrazak, Sheikh et~al.}]{abd2023large}
Alaa Abd-Alrazaq, Rawan AlSaad, Dari Alhuwail, Arfan Ahmed, Padraig~Mark Healy, Syed Latifi, Sarah Aziz, Rafat Damseh, Sadam~Alabed Alrazak, Javaid Sheikh, and 1 others. 2023.
\newblock Large language models in medical education: opportunities, challenges, and future directions.
\newblock \emph{JMIR Medical Education}, 9(1):e48291.

\bibitem[{Abdelrahman et~al.(2023)Abdelrahman, Wang, and Nunes}]{abdelrahman2023knowledge}
Ghodai Abdelrahman, Qing Wang, and Bernardo Nunes. 2023.
\newblock Knowledge tracing: A survey.
\newblock \emph{ACM Computing Surveys}, 55(11):1--37.

\bibitem[{Abu-Rasheed et~al.(2024{\natexlab{a}})Abu-Rasheed, Abdulsalam, Weber, and Fathi}]{abu2024supporting}
Hasan Abu-Rasheed, Mohamad~Hussam Abdulsalam, Christian Weber, and Madjid Fathi. 2024{\natexlab{a}}.
\newblock Supporting student decisions on learning recommendations: An llm-based chatbot with knowledge graph contextualization for conversational explainability and mentoring.
\newblock \emph{arXiv preprint arXiv:2401.08517}.

\bibitem[{Abu-Rasheed et~al.(2024{\natexlab{b}})Abu-Rasheed, Weber, and Fathi}]{abu2024knowledge}
Hasan Abu-Rasheed, Christian Weber, and Madjid Fathi. 2024{\natexlab{b}}.
\newblock Knowledge graphs as context sources for llm-based explanations of learning recommendations.
\newblock In \emph{2024 IEEE Global Engineering Education Conference (EDUCON)}, pages 1--5. IEEE.

\bibitem[{Adetayo et~al.(2024)Adetayo, Aborisade, and Sanni}]{adetayo2024microsoft}
Adebowale~Jeremy Adetayo, Mariam~Oyinda Aborisade, and Basheer~Abiodun Sanni. 2024.
\newblock Microsoft copilot and anthropic claude ai in education and library service.
\newblock \emph{Library Hi Tech News}.

\bibitem[{Adewumi et~al.(2023)Adewumi, Alkhaled, Buck, Hernandez, Brilioth, Kekung, Ragimov, and Barney}]{adewumi2023procot}
Tosin Adewumi, Lama Alkhaled, Claudia Buck, Sergio Hernandez, Saga Brilioth, Mkpe Kekung, Yelvin Ragimov, and Elisa Barney. 2023.
\newblock Procot: Stimulating critical thinking and writing of students through engagement with large language models (llms).
\newblock \emph{arXiv preprint arXiv:2312.09801}.

\bibitem[{Adewumi et~al.(2024)Adewumi, Alkhaled, Gurung, van Boven, and Pagliai}]{adewumi2024fairness}
Tosin Adewumi, Lama Alkhaled, Namrata Gurung, Goya van Boven, and Irene Pagliai. 2024.
\newblock Fairness and bias in multimodal ai: A survey.
\newblock \emph{arXiv preprint arXiv:2406.19097}.

\bibitem[{Aird et~al.(2024)Aird, Farastu, Sun, Stefancov{\'a}, All, Voida, Mattei, and Burke}]{aird2024dynamic}
Amanda Aird, Paresha Farastu, Joshua Sun, Elena Stefancov{\'a}, Cassidy All, Amy Voida, Nicholas Mattei, and Robin Burke. 2024.
\newblock Dynamic fairness-aware recommendation through multi-agent social choice.
\newblock \emph{ACM Transactions on Recommender Systems}, 3(2):1--35.

\bibitem[{Antoniades et~al.(2024)Antoniades, {\"O}rwall, Zhang, Xie, Goyal, and Wang}]{antoniades2024swe}
Antonis Antoniades, Albert {\"O}rwall, Kexun Zhang, Yuxi Xie, Anirudh Goyal, and William Wang. 2024.
\newblock Swe-search: Enhancing software agents with monte carlo tree search and iterative refinement.
\newblock \emph{arXiv preprint arXiv:2410.20285}.

\bibitem[{Ao et~al.(2024)Ao, Wang, Tian, Chen, Zhang, Lu, Wang, Li, and Wu}]{ao2024sdeval}
Junyi Ao, Yuancheng Wang, Xiaohai Tian, Dekun Chen, Jun Zhang, Lu~Lu, Yuxuan Wang, Haizhou Li, and Zhizheng Wu. 2024.
\newblock \href {https://arxiv.org/abs/2406.13340} {Sd-eval: A benchmark dataset for spoken dialogue understanding beyond words}.

\bibitem[{Asai et~al.(2024)Asai, He, Shao, Shi, Singh, Chang, Lo, Soldaini, Feldman, D'arcy et~al.}]{asai2024openscholar}
Akari Asai, Jacqueline He, Rulin Shao, Weijia Shi, Amanpreet Singh, Joseph~Chee Chang, Kyle Lo, Luca Soldaini, Sergey Feldman, Mike D'arcy, and 1 others. 2024.
\newblock Openscholar: Synthesizing scientific literature with retrieval-augmented lms.
\newblock \emph{arXiv preprint arXiv:2411.14199}.

\bibitem[{Baek et~al.(2024)Baek, Jauhar, Cucerzan, and Hwang}]{baek2024researchagent}
Jinheon Baek, Sujay~Kumar Jauhar, Silviu Cucerzan, and Sung~Ju Hwang. 2024.
\newblock Researchagent: Iterative research idea generation over scientific literature with large language models.
\newblock \emph{arXiv preprint arXiv:2404.07738}.

\bibitem[{Balan et~al.(2024)Balan, Dobrean, and Poetar}]{balan2024use}
Raluca Balan, Anca Dobrean, and Costina~R Poetar. 2024.
\newblock Use of automated conversational agents in improving young population mental health: a scoping review.
\newblock \emph{NPJ Digital Medicine}, 7(1):75.

\bibitem[{Barman et~al.(2025)Barman, Caron, Sullivan, de~Regt, de~Austri, Boon, F{\"a}rber, Fr{\"o}se, Hasibi, Ipp et~al.}]{barman2025large}
Kristian~G Barman, Sascha Caron, Emily Sullivan, Henk~W de~Regt, Roberto~Ruiz de~Austri, Mieke Boon, Michael F{\"a}rber, Stefan Fr{\"o}se, Faegheh Hasibi, Andreas Ipp, and 1 others. 2025.
\newblock Large physics models: Towards a collaborative approach with large language models and foundation models.
\newblock \emph{arXiv preprint arXiv:2501.05382}.

\bibitem[{Bassner et~al.(2024)Bassner, Frankford, and Krusche}]{bassner2024iris}
Patrick Bassner, Eduard Frankford, and Stephan Krusche. 2024.
\newblock Iris: An ai-driven virtual tutor for computer science education.
\newblock In \emph{Proceedings of the 2024 on Innovation and Technology in Computer Science Education V. 1}, pages 394--400.

\bibitem[{Beredo and Ong(2021)}]{beredo2021beyond}
Jackylyn Beredo and Ethel Ong. 2021.
\newblock Beyond the scene: A comparative analysis of two storytelling-based conversational agents.
\newblock In \emph{Proceedings of the Asian CHI Symposium 2021}, pages 189--195.

\bibitem[{Bewersdorff et~al.(2025)Bewersdorff, Hartmann, Hornberger, Se{\ss}ler, Bannert, Kasneci, Kasneci, Zhai, and Nerdel}]{bewersdorff2025taking}
Arne Bewersdorff, Christian Hartmann, Marie Hornberger, Kathrin Se{\ss}ler, Maria Bannert, Enkelejda Kasneci, Gjergji Kasneci, Xiaoming Zhai, and Claudia Nerdel. 2025.
\newblock Taking the next step with generative artificial intelligence: The transformative role of multimodal large language models in science education.
\newblock \emph{Learning and Individual Differences}, 118:102601.

\bibitem[{Bhattacharya et~al.(2024)Bhattacharya, Pal, Chatterjee, Lee, and Chakraborty}]{bhattacharya2024large}
Manojit Bhattacharya, Soumen Pal, Srijan Chatterjee, Sang-Soo Lee, and Chiranjib Chakraborty. 2024.
\newblock Large language model to multimodal large language model: A journey to shape the biological macromolecules to biological sciences and medicine.
\newblock \emph{Molecular Therapy-Nucleic Acids}, 35(3).

\bibitem[{Bhowmik et~al.(2024)Bhowmik, West, Barrett, Zhang, Dai, Sokolikj, Southerland, Yuan, and Ke}]{bhowmik2024evaluation}
Saptarshi Bhowmik, Luke West, Alex Barrett, Nuodi Zhang, Chih-Pu Dai, Zlatko Sokolikj, Sherry Southerland, Xin Yuan, and Fengfeng Ke. 2024.
\newblock Evaluation of an llm-powered student agent for teacher training.
\newblock In \emph{European Conference on Technology Enhanced Learning}, pages 68--74. Springer.

\bibitem[{Bogin et~al.(2024)Bogin, Yang, Gupta, Richardson, Bransom, Clark, Sabharwal, and Khot}]{bogin2024super}
Ben Bogin, Kejuan Yang, Shashank Gupta, Kyle Richardson, Erin Bransom, Peter Clark, Ashish Sabharwal, and Tushar Khot. 2024.
\newblock Super: Evaluating agents on setting up and executing tasks from research repositories.
\newblock \emph{arXiv preprint arXiv:2409.07440}.

\bibitem[{Bouzenia et~al.(2024)Bouzenia, Devanbu, and Pradel}]{bouzenia2024repairagent}
Islem Bouzenia, Premkumar Devanbu, and Michael Pradel. 2024.
\newblock Repairagent: An autonomous, llm-based agent for program repair.
\newblock \emph{arXiv preprint arXiv:2403.17134}.

\bibitem[{Brown and Cruz~Castro(2025)}]{brown2025coordinate}
Cameron Brown and Laura Cruz~Castro. 2025.
\newblock Coordinate: A virtual classroom management tool for large computer science courses using discord.
\newblock In \emph{Proceedings of the 56th ACM Technical Symposium on Computer Science Education V. 1}, pages 165--171.

\bibitem[{Cai et~al.(2024)Cai, Wang, Wang, de~Melo, Zhang, Wang, and He}]{cai2024medbench}
Yan Cai, Linlin Wang, Ye~Wang, Gerard de~Melo, Ya~Zhang, Yanfeng Wang, and Liang He. 2024.
\newblock Medbench: A large-scale chinese benchmark for evaluating medical large language models.
\newblock In \emph{Proceedings of the AAAI Conference on Artificial Intelligence}, volume~38, pages 17709--17717.

\bibitem[{Cao et~al.(2024)Cao, Lei, Wu, Chen, Fu, Gao, Xiong, Zhang, Hu, Mao et~al.}]{cao2024spider2}
Ruisheng Cao, Fangyu Lei, Haoyuan Wu, Jixuan Chen, Yeqiao Fu, Hongcheng Gao, Xinzhuang Xiong, Hanchong Zhang, Wenjing Hu, Yuchen Mao, and 1 others. 2024.
\newblock Spider2-v: How far are multimodal agents from automating data science and engineering workflows?
\newblock \emph{Advances in Neural Information Processing Systems}, 37:107703--107744.

\bibitem[{Cassell(2022)}]{cassell2022socially}
Justine Cassell. 2022.
\newblock Socially interactive agents as peers.
\newblock In \emph{The Handbook on Socially Interactive Agents: 20 years of Research on Embodied Conversational Agents, Intelligent Virtual Agents, and Social Robotics Volume 2: Interactivity, Platforms, Application}, pages 331--366.

\bibitem[{Chang et~al.(2015)Chang, Hsu, Chen et~al.}]{chang2015modeling}
Haw-Shiuan Chang, Hwai-Jung Hsu, Kuan-Ta Chen, and 1 others. 2015.
\newblock Modeling exercise relationships in e-learning: A unified approach.
\newblock In \emph{EDM}, pages 532--535.

\bibitem[{Chen et~al.(2024{\natexlab{a}})Chen, Wei, Le, and Zhang}]{chen2024learning}
Angxuan Chen, Yuang Wei, Huixiao Le, and Yan Zhang. 2024{\natexlab{a}}.
\newblock Learning-by-teaching with chatgpt: The effect of teachable chatgpt agent on programming education.
\newblock \emph{arXiv preprint arXiv:2412.15226}.

\bibitem[{Chen and Leitch(2024)}]{chen2024llms}
Celia Chen and Alex Leitch. 2024.
\newblock Llms as academic reading companions: Extending hci through synthetic personae.
\newblock \emph{arXiv preprint arXiv:2403.19506}.

\bibitem[{Chen et~al.(2024{\natexlab{b}})Chen, Fan, Gong, Xie, Li, Liu, Li, Qu, Ni, and Yang}]{chen2024agentcourt}
Guhong Chen, Liyang Fan, Zihan Gong, Nan Xie, Zixuan Li, Ziqiang Liu, Chengming Li, Qiang Qu, Shiwen Ni, and Min Yang. 2024{\natexlab{b}}.
\newblock Agentcourt: Simulating court with adversarial evolvable lawyer agents.
\newblock \emph{arXiv preprint arXiv:2408.08089}.

\bibitem[{Chen et~al.(2023)Chen, Pasunuru, Weston, and Celikyilmaz}]{chen2023walking}
Howard Chen, Ramakanth Pasunuru, Jason Weston, and Asli Celikyilmaz. 2023.
\newblock Walking down the memory maze: Beyond context limit through interactive reading.
\newblock \emph{arXiv preprint arXiv:2310.05029}.

\bibitem[{Chen et~al.(2025{\natexlab{a}})Chen, Ding, Yu, Huang, Yang, and Zhang}]{chen2025scitoolagent}
Huajun Chen, Keyan Ding, Jing Yu, Junjie Huang, Yuchen Yang, and Qiang Zhang. 2025{\natexlab{a}}.
\newblock Scitoolagent: A knowledge graph-driven scientific agent for multi-tool integration.

\bibitem[{Chen et~al.()Chen, Wang, Xu, Yuan, Zhang, Shi, Xie, Li, Yang, Zhu et~al.}]{chenpersona}
Jiangjie Chen, Xintao Wang, Rui Xu, Siyu Yuan, Yikai Zhang, Wei Shi, Jian Xie, Shuang Li, Ruihan Yang, Tinghui Zhu, and 1 others.
\newblock From persona to personalization: A survey on role-playing language agents.
\newblock \emph{Transactions on Machine Learning Research}.

\bibitem[{Chen et~al.(2025{\natexlab{b}})Chen, Deng, Zheng, Yan, Liu, Wu, Jiang, Liu, and Hu}]{chen2025safeeraser}
Junkai Chen, Zhijie Deng, Kening Zheng, Yibo Yan, Shuliang Liu, PeiJun Wu, Peijie Jiang, Jia Liu, and Xuming Hu. 2025{\natexlab{b}}.
\newblock Safeeraser: Enhancing safety in multimodal large language models through multimodal machine unlearning.
\newblock \emph{arXiv preprint arXiv:2502.12520}.

\bibitem[{Chen et~al.(2020)Chen, Chen, and Lin}]{chen2020artificial}
Lijia Chen, Pingping Chen, and Zhijian Lin. 2020.
\newblock Artificial intelligence in education: A review.
\newblock \emph{Ieee Access}, 8:75264--75278.

\bibitem[{Chen et~al.(2024{\natexlab{c}})Chen, Ding, Zheng, Liu, Sun, and Zhou}]{chen2024empowering}
Yulin Chen, Ning Ding, Hai-Tao Zheng, Zhiyuan Liu, Maosong Sun, and Bowen Zhou. 2024{\natexlab{c}}.
\newblock Empowering private tutoring by chaining large language models.
\newblock In \emph{Proceedings of the 33rd ACM International Conference on Information and Knowledge Management}, pages 354--364.

\bibitem[{Chen et~al.(2024{\natexlab{d}})Chen, Chen, Ning, Zhang, Wang, Yu, Li, Liao, Wei, Lu et~al.}]{chen2024scienceagentbench}
Ziru Chen, Shijie Chen, Yuting Ning, Qianheng Zhang, Boshi Wang, Botao Yu, Yifei Li, Zeyi Liao, Chen Wei, Zitong Lu, and 1 others. 2024{\natexlab{d}}.
\newblock Scienceagentbench: Toward rigorous assessment of language agents for data-driven scientific discovery.
\newblock \emph{arXiv preprint arXiv:2410.05080}.

\bibitem[{Cheng et~al.(2024)Cheng, Huang, Ko, Li, Peng, Xu, and Zhang}]{cheng2024towards}
Shanbo Cheng, Zhichao Huang, Tom Ko, Hang Li, Ningxin Peng, Lu~Xu, and Qini Zhang. 2024.
\newblock Towards achieving human parity on end-to-end simultaneous speech translation via llm agent.
\newblock \emph{arXiv preprint arXiv:2407.21646}.

\bibitem[{Cherakara et~al.(2023)Cherakara, Varghese, Shabana, Nelson, Karukayil, Kulothungan, Afil~Farhan, Nesset, Moujahid, Dinkar, Rieser, and Lemon}]{cherakara-etal-2023-furchat}
Neeraj Cherakara, Finny Varghese, Sheena Shabana, Nivan Nelson, Abhiram Karukayil, Rohith Kulothungan, Mohammed Afil~Farhan, Birthe Nesset, Meriam Moujahid, Tanvi Dinkar, Verena Rieser, and Oliver Lemon. 2023.
\newblock \href {https://doi.org/10.18653/v1/2023.sigdial-1.55} {{F}ur{C}hat: An embodied conversational agent using {LLM}s, combining open and closed-domain dialogue with facial expressions}.
\newblock In \emph{Proceedings of the 24th Annual Meeting of the Special Interest Group on Discourse and Dialogue}, pages 588--592, Prague, Czechia. Association for Computational Linguistics.

\bibitem[{Chu et~al.(2025)Chu, Xie, Wang, Wang, and Wen}]{chu2025uniedu}
Zhendong Chu, Jian Xie, Shen Wang, Zichao Wang, and Qingsong Wen. 2025.
\newblock Uniedu: A unified language and vision assistant for education applications.
\newblock \emph{arXiv preprint arXiv:2503.20701}.

\bibitem[{da~Silva et~al.(2024)da~Silva, Fonseca, Labidi, and Pacheco}]{da2024mitigation}
Wildemarkes de~Almeida da~Silva, Luis Carlos~Costa Fonseca, Sofiane Labidi, and Jos{\'e} Chrystian~Lima Pacheco. 2024.
\newblock Mitigation of hallucinations in language models in education: A new approach of comparative and cross-verification.
\newblock In \emph{2024 IEEE International Conference on Advanced Learning Technologies (ICALT)}, pages 207--209. IEEE.

\bibitem[{Ding et~al.(2025)Ding, Tao, Pang, Wei, Gao, Ding, Shen, and Chen}]{ding2025toolcoder}
Hanxing Ding, Shuchang Tao, Liang Pang, Zihao Wei, Jinyang Gao, Bolin Ding, Huawei Shen, and Xueqi Chen. 2025.
\newblock Toolcoder: A systematic code-empowered tool learning framework for large language models.
\newblock \emph{arXiv preprint arXiv:2502.11404}.

\bibitem[{Du et~al.(2024)Du, Wang, Zhao, Deng, Liu, Lou, Zou, Venkit, Zhang, Srinath et~al.}]{du2024llms}
Jiangshu Du, Yibo Wang, Wenting Zhao, Zhongfen Deng, Shuaiqi Liu, Renze Lou, Henry Zou, Pranav~Narayanan Venkit, Nan Zhang, Mukund Srinath, and 1 others. 2024.
\newblock Llms assist nlp researchers: Critique paper (meta-) reviewing.
\newblock In \emph{Proceedings of the 2024 Conference on Empirical Methods in Natural Language Processing}, pages 5081--5099.

\bibitem[{Est{\'e}vez-Ayres et~al.(2024)Est{\'e}vez-Ayres, Callejo, Hombrados-Herrera, Alario-Hoyos, and Delgado~Kloos}]{estevez2024evaluation}
Iria Est{\'e}vez-Ayres, Patricia Callejo, Miguel~{\'A}ngel Hombrados-Herrera, Carlos Alario-Hoyos, and Carlos Delgado~Kloos. 2024.
\newblock Evaluation of llm tools for feedback generation in a course on concurrent programming.
\newblock \emph{International Journal of Artificial Intelligence in Education}, pages 1--17.

\bibitem[{Fang et~al.(2025)Fang, Qiang, Ouyang, Zhu, Yuan, and Li}]{fang-etal-2025-collaborative}
Dengzhao Fang, Jipeng Qiang, Xiaoye Ouyang, Yi~Zhu, Yunhao Yuan, and Yun Li. 2025.
\newblock \href {https://aclanthology.org/2025.coling-main.60/} {Collaborative document simplification using multi-agent systems}.
\newblock In \emph{Proceedings of the 31st International Conference on Computational Linguistics}, pages 897--912, Abu Dhabi, UAE. Association for Computational Linguistics.

\bibitem[{Fei et~al.(2023)Fei, Shen, Zhu, Zhou, Han, Zhang, Chen, Shen, and Ge}]{fei2023lawbench}
Zhiwei Fei, Xiaoyu Shen, Dawei Zhu, Fengzhe Zhou, Zhuo Han, Songyang Zhang, Kai Chen, Zongwen Shen, and Jidong Ge. 2023.
\newblock Lawbench: Benchmarking legal knowledge of large language models.
\newblock \emph{arXiv preprint arXiv:2309.16289}.

\bibitem[{Feng et~al.()Feng, Huang, Liu, Jiang, and Yan}]{fengphyspde}
Mingquan Feng, Yixin Huang, Yizhou Liu, Bofang Jiang, and Junchi Yan.
\newblock Physpde: Rethinking pde discovery and a physical hypothesis selection benchmark.
\newblock In \emph{The Thirteenth International Conference on Learning Representations}.

\bibitem[{Feng et~al.(2009)Feng, Heffernan, and Koedinger}]{feng2009addressing}
Mingyu Feng, Neil Heffernan, and Kenneth Koedinger. 2009.
\newblock Addressing the assessment challenge with an online system that tutors as it assesses.
\newblock \emph{User modeling and user-adapted interaction}, 19:243--266.

\bibitem[{Fourney et~al.(2024)Fourney, Bansal, Mozannar, Tan, Salinas, Niedtner, Proebsting, Bassman, Gerrits, Alber et~al.}]{fourney2024magentic}
Adam Fourney, Gagan Bansal, Hussein Mozannar, Cheng Tan, Eduardo Salinas, Friederike Niedtner, Grace Proebsting, Griffin Bassman, Jack Gerrits, Jacob Alber, and 1 others. 2024.
\newblock Magentic-one: A generalist multi-agent system for solving complex tasks.
\newblock \emph{arXiv preprint arXiv:2411.04468}.

\bibitem[{Frankford et~al.(2024)Frankford, H{\"o}hn, Sauerwein, and Breu}]{frankford2024survey}
Eduard Frankford, Ingo H{\"o}hn, Clemens Sauerwein, and Ruth Breu. 2024.
\newblock A survey study on the state of the art of programming exercise generation using large language models.
\newblock In \emph{2024 36th International Conference on Software Engineering Education and Training (CSEE\&T)}, pages 1--5. IEEE.

\bibitem[{Gan et~al.(2024)Gan, Yang, Ma, He, Zeng, Wang, Li, Zhou, Li, Wang et~al.}]{gan2024navigating}
Yuyou Gan, Yong Yang, Zhe Ma, Ping He, Rui Zeng, Yiming Wang, Qingming Li, Chunyi Zhou, Songze Li, Ting Wang, and 1 others. 2024.
\newblock Navigating the risks: A survey of security, privacy, and ethics threats in llm-based agents.
\newblock \emph{arXiv preprint arXiv:2411.09523}.

\bibitem[{Gao et~al.(2024)Gao, Fang, Huang, Giunchiglia, Noori, Schwarz, Ektefaie, Kondic, and Zitnik}]{gao2024empowering}
Shanghua Gao, Ada Fang, Yepeng Huang, Valentina Giunchiglia, Ayush Noori, Jonathan~Richard Schwarz, Yasha Ektefaie, Jovana Kondic, and Marinka Zitnik. 2024.
\newblock Empowering biomedical discovery with ai agents.
\newblock \emph{Cell}, 187(22):6125--6151.

\bibitem[{Gao et~al.(2021)Gao, Liu, Huang, Yin, Bi, Wang, Ma, Wang, and Su}]{gao2021rcd}
Weibo Gao, Qi~Liu, Zhenya Huang, Yu~Yin, Haoyang Bi, Mu-Chun Wang, Jianhui Ma, Shijin Wang, and Yu~Su. 2021.
\newblock Rcd: Relation map driven cognitive diagnosis for intelligent education systems.
\newblock In \emph{Proceedings of the 44th international ACM SIGIR conference on research and development in information retrieval}, pages 501--510.

\bibitem[{Gao et~al.(2023)Gao, Xiong, Gao, Jia, Pan, Bi, Dai, Sun, Wang, and Wang}]{gao2023retrieval}
Yunfan Gao, Yun Xiong, Xinyu Gao, Kangxiang Jia, Jinliu Pan, Yuxi Bi, Yi~Dai, Jiawei Sun, Haofen Wang, and Haofen Wang. 2023.
\newblock Retrieval-augmented generation for large language models: A survey.
\newblock \emph{arXiv preprint arXiv:2312.10997}, 2.

\bibitem[{Ghafarollahi and Buehler(2024{\natexlab{a}})}]{ghafarollahi2024protagents}
Alireza Ghafarollahi and Markus~J Buehler. 2024{\natexlab{a}}.
\newblock Protagents: protein discovery via large language model multi-agent collaborations combining physics and machine learning.
\newblock \emph{Digital Discovery}, 3(7):1389--1409.

\bibitem[{Ghafarollahi and Buehler(2024{\natexlab{b}})}]{ghafarollahi2024sciagents}
Alireza Ghafarollahi and Markus~J Buehler. 2024{\natexlab{b}}.
\newblock Sciagents: Automating scientific discovery through bioinspired multi-agent intelligent graph reasoning.
\newblock \emph{Advanced Materials}, page 2413523.

\bibitem[{Gou et~al.(2023)Gou, Shao, Gong, Shen, Yang, Huang, Duan, and Chen}]{gou2023tora}
Zhibin Gou, Zhihong Shao, Yeyun Gong, Yelong Shen, Yujiu Yang, Minlie Huang, Nan Duan, and Weizhu Chen. 2023.
\newblock Tora: A tool-integrated reasoning agent for mathematical problem solving.
\newblock \emph{arXiv preprint arXiv:2309.17452}.

\bibitem[{Grosnit et~al.(2024)Grosnit, Maraval, Doran, Paolo, Thomas, Beevi, Gonzalez, Khandelwal, Iacobacci, Benechehab et~al.}]{grosnit2024large}
Antoine Grosnit, Alexandre Maraval, James Doran, Giuseppe Paolo, Albert Thomas, Refinath Shahul Hameed~Nabeezath Beevi, Jonas Gonzalez, Khyati Khandelwal, Ignacio Iacobacci, Abdelhakim Benechehab, and 1 others. 2024.
\newblock Large language models orchestrating structured reasoning achieve kaggle grandmaster level.
\newblock \emph{arXiv preprint arXiv:2411.03562}.

\bibitem[{Gu et~al.(2024)Gu, Shang, Jiang, Kuang, Lin, Lyu, Mao, Pan, Wu, Yu et~al.}]{gu2024blade}
Ken Gu, Ruoxi Shang, Ruien Jiang, Keying Kuang, Richard-John Lin, Donghe Lyu, Yue Mao, Youran Pan, Teng Wu, Jiaqian Yu, and 1 others. 2024.
\newblock Blade: Benchmarking language model agents for data-driven science.
\newblock \emph{arXiv preprint arXiv:2408.09667}.

\bibitem[{Guha et~al.(2023)Guha, Nyarko, Ho, R{\'e}, Chilton, Chohlas-Wood, Peters, Waldon, Rockmore, Zambrano et~al.}]{guha2023legalbench}
Neel Guha, Julian Nyarko, Daniel Ho, Christopher R{\'e}, Adam Chilton, Alex Chohlas-Wood, Austin Peters, Brandon Waldon, Daniel Rockmore, Diego Zambrano, and 1 others. 2023.
\newblock Legalbench: A collaboratively built benchmark for measuring legal reasoning in large language models.
\newblock \emph{Advances in Neural Information Processing Systems}, 36:44123--44279.

\bibitem[{Guo et~al.(2024{\natexlab{a}})Guo, Zhang, Ma, Zhang, and Feng}]{guo2024agent}
Shoutao Guo, Shaolei Zhang, Zhengrui Ma, Min Zhang, and Yang Feng. 2024{\natexlab{a}}.
\newblock Agent-simt: Agent-assisted simultaneous machine translation with large language models.
\newblock \emph{arXiv preprint arXiv:2406.06910}.

\bibitem[{Guo et~al.(2024{\natexlab{b}})Guo, Zhang, Ma, Zhang, and Feng}]{guo2024sillm}
Shoutao Guo, Shaolei Zhang, Zhengrui Ma, Min Zhang, and Yang Feng. 2024{\natexlab{b}}.
\newblock Sillm: Large language models for simultaneous machine translation.
\newblock \emph{arXiv preprint arXiv:2402.13036}.

\bibitem[{Guo et~al.(2024{\natexlab{c}})Guo, Latif, Zhou, Huang, and Zhai}]{guo2024using}
Shuchen Guo, Ehsan Latif, Yifan Zhou, Xuan Huang, and Xiaoming Zhai. 2024{\natexlab{c}}.
\newblock Using generative ai and multi-agents to provide automatic feedback.
\newblock \emph{arXiv preprint arXiv:2411.07407}.

\bibitem[{Guo et~al.(2023)Guo, Nan, Liang, Guo, Chawla, Wiest, Zhang et~al.}]{guo2023can}
Taicheng Guo, Bozhao Nan, Zhenwen Liang, Zhichun Guo, Nitesh Chawla, Olaf Wiest, Xiangliang Zhang, and 1 others. 2023.
\newblock What can large language models do in chemistry? a comprehensive benchmark on eight tasks.
\newblock \emph{Advances in Neural Information Processing Systems}, 36:59662--59688.

\bibitem[{Han et~al.(2023)Han, Yoo, Myung, Kim, Lim, Kim, Lee, Hong, Kim, Ahn et~al.}]{han2023fabric}
Jieun Han, Haneul Yoo, Junho Myung, Minsun Kim, Hyunseung Lim, Yoonsu Kim, Tak~Yeon Lee, Hwajung Hong, Juho Kim, So-Yeon Ahn, and 1 others. 2023.
\newblock Fabric: Automated scoring and feedback generation for essays.
\newblock \emph{arXiv preprint arXiv:2310.05191}.

\bibitem[{He et~al.(2024{\natexlab{a}})He, Zhu, Ye, Liu, Zhou, and Yu}]{he2024emerged}
Feng He, Tianqing Zhu, Dayong Ye, Bo~Liu, Wanlei Zhou, and Philip~S Yu. 2024{\natexlab{a}}.
\newblock The emerged security and privacy of llm agent: A survey with case studies.
\newblock \emph{arXiv preprint arXiv:2407.19354}.

\bibitem[{He et~al.(2023)He, Wang, Yan, Liu, Chang, Gentili, McAuley, and Hsu}]{he-etal-2023-medeval}
Zexue He, Yu~Wang, An~Yan, Yao Liu, Eric Chang, Amilcare Gentili, Julian McAuley, and Chun-Nan Hsu. 2023.
\newblock \href {https://doi.org/10.18653/v1/2023.emnlp-main.540} {{M}ed{E}val: A multi-level, multi-task, and multi-domain medical benchmark for language model evaluation}.
\newblock In \emph{Proceedings of the 2023 Conference on Empirical Methods in Natural Language Processing}, pages 8725--8744, Singapore. Association for Computational Linguistics.

\bibitem[{He et~al.(2024{\natexlab{b}})He, Cao, Wang, Jin, Chen, Xu, Li, Jiang, Liu, and Zhao}]{he2024agentscourt}
Zhitao He, Pengfei Cao, Chenhao Wang, Zhuoran Jin, Yubo Chen, Jiexin Xu, Huaijun Li, Xiaojian Jiang, Kang Liu, and Jun Zhao. 2024{\natexlab{b}}.
\newblock Agentscourt: Building judicial decision-making agents with court debate simulation and legal knowledge augmentation.
\newblock \emph{arXiv preprint arXiv:2403.02959}.

\bibitem[{Ho et~al.(2024)Ho, Ly, and Nguyen}]{ho2024mitigating}
Huu-Tuong Ho, Duc-Tin Ly, and Luong~Vuong Nguyen. 2024.
\newblock Mitigating hallucinations in large language models for educational application.
\newblock In \emph{2024 IEEE International Conference on Consumer Electronics-Asia (ICCE-Asia)}, pages 1--4. IEEE.

\bibitem[{Hong et~al.(2024)Hong, Cai, Du, Feng, Liu, and Fan}]{hong2024my}
Shengxin Hong, Chang Cai, Sixuan Du, Haiyue Feng, Siyuan Liu, and Xiuyi Fan. 2024.
\newblock " my grade is wrong!": A contestable ai framework for interactive feedback in evaluating student essays.
\newblock \emph{arXiv preprint arXiv:2409.07453}.

\bibitem[{Hou et~al.(2024)Hou, Ao, Wu, Kong, Zheng, Tang, Li, Hu, Xu, Ni et~al.}]{hou2024eval}
Jinchang Hou, Chang Ao, Haihong Wu, Xiangtao Kong, Zhigang Zheng, Daijia Tang, Chengming Li, Xiping Hu, Ruifeng Xu, Shiwen Ni, and 1 others. 2024.
\newblock E-eval: a comprehensive chinese k-12 education evaluation benchmark for large language models.
\newblock \emph{arXiv preprint arXiv:2401.15927}.

\bibitem[{Hu et~al.(2025{\natexlab{a}})Hu, Zhu, Pei, and Gu}]{hu2025exploring}
Bihao Hu, Jiayi Zhu, Yiying Pei, and Xiaoqing Gu. 2025{\natexlab{a}}.
\newblock Exploring the potential of llm to enhance teaching plans through teaching simulation.
\newblock \emph{npj Science of Learning}, 10(1):7.

\bibitem[{Hu and Wang(2024)}]{hu2024foke}
Silan Hu and Xiaoning Wang. 2024.
\newblock Foke: A personalized and explainable education framework integrating foundation models, knowledge graphs, and prompt engineering.
\newblock In \emph{China National Conference on Big Data and Social Computing}, pages 399--411. Springer.

\bibitem[{Hu et~al.(2024)Hu, Li, Lu, Shao, He, Qiao, and Luo}]{hu2024omnimedvqa}
Yutao Hu, Tianbin Li, Quanfeng Lu, Wenqi Shao, Junjun He, Yu~Qiao, and Ping Luo. 2024.
\newblock Omnimedvqa: A new large-scale comprehensive evaluation benchmark for medical lvlm.
\newblock In \emph{Proceedings of the IEEE/CVF Conference on Computer Vision and Pattern Recognition}, pages 22170--22183.

\bibitem[{Hu et~al.(2025{\natexlab{b}})Hu, Chan, Li, and Yin}]{hu-etal-2025-debate}
Zhe Hu, Hou~Pong Chan, Jing Li, and Yu~Yin. 2025{\natexlab{b}}.
\newblock \href {https://aclanthology.org/2025.coling-main.314/} {Debate-to-write: A persona-driven multi-agent framework for diverse argument generation}.
\newblock In \emph{Proceedings of the 31st International Conference on Computational Linguistics}, pages 4689--4703, Abu Dhabi, UAE. Association for Computational Linguistics.

\bibitem[{Hua et~al.(2024)Hua, Yang, Jin, Li, Cheng, Tang, and Zhang}]{hua2024trustagent}
Wenyue Hua, Xianjun Yang, Mingyu Jin, Zelong Li, Wei Cheng, Ruixiang Tang, and Yongfeng Zhang. 2024.
\newblock Trustagent: Towards safe and trustworthy llm-based agents through agent constitution.
\newblock In \emph{Trustworthy Multi-modal Foundation Models and AI Agents (TiFA)}.

\bibitem[{Huang et~al.(2023{\natexlab{a}})Huang, Zhang, Luck, Bu, Qing, and Cui}]{huang2023agentcoder}
Dong Huang, Jie~M Zhang, Michael Luck, Qingwen Bu, Yuhao Qing, and Heming Cui. 2023{\natexlab{a}}.
\newblock Agentcoder: Multi-agent-based code generation with iterative testing and optimisation.
\newblock \emph{arXiv preprint arXiv:2312.13010}.

\bibitem[{Huang et~al.(2023{\natexlab{b}})Huang, Ye, Zhou, Li, Li, Zhou, and Zheng}]{huang2023frustratingly}
Haojing Huang, Jingheng Ye, Qingyu Zhou, Yinghui Li, Yangning Li, Feng Zhou, and Hai-Tao Zheng. 2023{\natexlab{b}}.
\newblock A frustratingly easy plug-and-play detection-and-reasoning module for chinese spelling check.
\newblock \emph{arXiv preprint arXiv:2310.09119}.

\bibitem[{Huang et~al.(2024{\natexlab{a}})Huang, Shi, Lei, Hu, and Cai}]{huang2024protchat}
Huazhen Huang, Xianguo Shi, Hongyang Lei, Fan Hu, and Yunpeng Cai. 2024{\natexlab{a}}.
\newblock Protchat: An ai multi-agent for automated protein analysis leveraging gpt-4 and protein language model.
\newblock \emph{Journal of Chemical Information and Modeling}, 65(1):62--70.

\bibitem[{Huang(2023)}]{huang2023ethics}
Lan Huang. 2023.
\newblock Ethics of artificial intelligence in education: Student privacy and data protection.
\newblock \emph{Science Insights Education Frontiers}, 16(2):2577--2587.

\bibitem[{Huang et~al.(2023{\natexlab{c}})Huang, Vora, Liang, and Leskovec}]{huang2023mlagentbench}
Qian Huang, Jian Vora, Percy Liang, and Jure Leskovec. 2023{\natexlab{c}}.
\newblock Mlagentbench: Evaluating language agents on machine learning experimentation.
\newblock \emph{arXiv preprint arXiv:2310.03302}.

\bibitem[{Huang et~al.(2024{\natexlab{b}})Huang, Liu, Chen, Wang, Wang, Lian, Wang, Tang, and Chen}]{huang2024understanding}
Xu~Huang, Weiwen Liu, Xiaolong Chen, Xingmei Wang, Hao Wang, Defu Lian, Yasheng Wang, Ruiming Tang, and Enhong Chen. 2024{\natexlab{b}}.
\newblock Understanding the planning of llm agents: A survey.
\newblock \emph{arXiv preprint arXiv:2402.02716}.

\bibitem[{Huo et~al.(2025)Huo, Yan, Zheng, Lyu, Zou, Wei, and Hu}]{huo2025mmunlearner}
Jiahao Huo, Yibo Yan, Xu~Zheng, Yuanhuiyi Lyu, Xin Zou, Zhihua Wei, and Xuming Hu. 2025.
\newblock Mmunlearner: Reformulating multimodal machine unlearning in the era of multimodal large language models.
\newblock \emph{arXiv preprint arXiv:2502.11051}.

\bibitem[{Islam et~al.(2024)Islam, Ali, and Parvez}]{islam2024mapcoder}
Md~Ashraful Islam, Mohammed~Eunus Ali, and Md~Rizwan Parvez. 2024.
\newblock Mapcoder: Multi-agent code generation for competitive problem solving.
\newblock \emph{arXiv preprint arXiv:2405.11403}.

\bibitem[{Ismail(2025)}]{ismail2025protecting}
Islam~Asim Ismail. 2025.
\newblock Protecting privacy in ai-enhanced education: A comprehensive examination of data privacy concerns and solutions in ai-based learning.
\newblock \emph{Impacts of Generative AI on the Future of Research and Education}, pages 117--142.

\bibitem[{Jablonka et~al.(2024)Jablonka, Schwaller, Ortega-Guerrero, and Smit}]{jablonka2024leveraging}
Kevin~Maik Jablonka, Philippe Schwaller, Andres Ortega-Guerrero, and Berend Smit. 2024.
\newblock Leveraging large language models for predictive chemistry.
\newblock \emph{Nature Machine Intelligence}, 6(2):161--169.

\bibitem[{Jansen et~al.(2024)Jansen, C{\^o}t{\'e}, Khot, Bransom, Dalvi~Mishra, Majumder, Tafjord, and Clark}]{jansen2024discoveryworld}
Peter Jansen, Marc-Alexandre C{\^o}t{\'e}, Tushar Khot, Erin Bransom, Bhavana Dalvi~Mishra, Bodhisattwa~Prasad Majumder, Oyvind Tafjord, and Peter Clark. 2024.
\newblock Discoveryworld: A virtual environment for developing and evaluating automated scientific discovery agents.
\newblock \emph{Advances in Neural Information Processing Systems}, 37:10088--10116.

\bibitem[{Jho(2024)}]{jho2024leveraging}
Hunkoog Jho. 2024.
\newblock Leveraging generative ai in physics education: Addressing hallucination issues in large language models.

\bibitem[{Jiang and Yang(2024)}]{jiang2024agents}
Cong Jiang and Xiaolei Yang. 2024.
\newblock Agents on the bench: Large language model based multi agent framework for trustworthy digital justice.
\newblock \emph{arXiv preprint arXiv:2412.18697}.

\bibitem[{Jiang and Jiang(2024)}]{jiang2024beyond}
Zhoumingju Jiang and Mengjun Jiang. 2024.
\newblock Beyond answers: Large language model-powered tutoring system in physics education for deep learning and precise understanding.
\newblock \emph{arXiv preprint arXiv:2406.10934}.

\bibitem[{Jin et~al.(2021)Jin, Pan, Oufattole, Weng, Fang, and Szolovits}]{jin2021disease}
Di~Jin, Eileen Pan, Nassim Oufattole, Wei-Hung Weng, Hanyi Fang, and Peter Szolovits. 2021.
\newblock What disease does this patient have? a large-scale open domain question answering dataset from medical exams.
\newblock \emph{Applied Sciences}, 11(14):6421.

\bibitem[{Jin et~al.(2024{\natexlab{a}})Jin, Lee, Shin, and Kim}]{jin2024teach}
Hyoungwook Jin, Seonghee Lee, Hyungyu Shin, and Juho Kim. 2024{\natexlab{a}}.
\newblock Teach ai how to code: Using large language models as teachable agents for programming education.
\newblock In \emph{Proceedings of the 2024 CHI Conference on Human Factors in Computing Systems}, pages 1--28.

\bibitem[{Jin et~al.(2024{\natexlab{b}})Jin, Yoo, Park, Lee, Wang, and Kim}]{jin2024teachtune}
Hyoungwook Jin, Minju Yoo, Jeongeon Park, Yokyung Lee, Xu~Wang, and Juho Kim. 2024{\natexlab{b}}.
\newblock Teachtune: Reviewing pedagogical agents against diverse student profiles with simulated students.
\newblock \emph{arXiv preprint arXiv:2410.04078}.

\bibitem[{Jinxin et~al.(2023)Jinxin, Jiabao, Yilei, Xingjiao, Jiawen, and Liang}]{jinxin2023cgmi}
Shi Jinxin, Zhao Jiabao, Wang Yilei, Wu~Xingjiao, Li~Jiawen, and He~Liang. 2023.
\newblock Cgmi: Configurable general multi-agent interaction framework.
\newblock \emph{arXiv preprint arXiv:2308.12503}.

\bibitem[{Karabacak and Margetis(2023)}]{karabacak2023embracing}
Mert Karabacak and Konstantinos Margetis. 2023.
\newblock Embracing large language models for medical applications: opportunities and challenges.
\newblock \emph{Cureus}, 15(5).

\bibitem[{Khan and Ghosh(2021)}]{khan2021student}
Anupam Khan and Soumya~K Ghosh. 2021.
\newblock Student performance analysis and prediction in classroom learning: A review of educational data mining studies.
\newblock \emph{Education and information technologies}, 26(1):205--240.

\bibitem[{Khan(2024)}]{khan2024ethical}
Wajahat~Naseeb Khan. 2024.
\newblock Ethical challenges of ai in education: Balancing innovation with data privacy.
\newblock \emph{Journal of AI in Education: Innovations, Opportunities, Challenges, and Future Directions}, 1(1):1--13.

\bibitem[{Koh et~al.(2024)Koh, Lo, Jang, Duvvur, Lim, Huang, Neubig, Zhou, Salakhutdinov, and Fried}]{koh2024visualwebarena}
Jing~Yu Koh, Robert Lo, Lawrence Jang, Vikram Duvvur, Ming~Chong Lim, Po-Yu Huang, Graham Neubig, Shuyan Zhou, Ruslan Salakhutdinov, and Daniel Fried. 2024.
\newblock Visualwebarena: Evaluating multimodal agents on realistic visual web tasks.
\newblock \emph{arXiv preprint arXiv:2401.13649}.

\bibitem[{Kon et~al.(2025)Kon, Liu, Ding, Qiu, Yang, Huang, Srinivasa, Lee, Chowdhury, and Chen}]{kon2025curie}
Patrick Tser~Jern Kon, Jiachen Liu, Qiuyi Ding, Yiming Qiu, Zhenning Yang, Yibo Huang, Jayanth Srinivasa, Myungjin Lee, Mosharaf Chowdhury, and Ang Chen. 2025.
\newblock Curie: Toward rigorous and automated scientific experimentation with ai agents.
\newblock \emph{arXiv preprint arXiv:2502.16069}.

\bibitem[{Kortemeyer(2023)}]{kortemeyer2023could}
Gerd Kortemeyer. 2023.
\newblock Could an artificial-intelligence agent pass an introductory physics course?
\newblock \emph{Physical Review Physics Education Research}, 19(1):010132.

\bibitem[{Kosar et~al.(2024)Kosar, Ostoji{\'c}, Liu, and Mernik}]{kosar2024computer}
Toma{\v{z}} Kosar, Dragana Ostoji{\'c}, Yu~David Liu, and Marjan Mernik. 2024.
\newblock Computer science education in chatgpt era: Experiences from an experiment in a programming course for novice programmers.
\newblock \emph{Mathematics}, 12(5):629.

\bibitem[{Koshkin et~al.(2024)Koshkin, Sudoh, and Nakamura}]{koshkin2024transllama}
Roman Koshkin, Katsuhito Sudoh, and Satoshi Nakamura. 2024.
\newblock Transllama: Llm-based simultaneous translation system.
\newblock \emph{arXiv preprint arXiv:2402.04636}.

\bibitem[{Krupp et~al.(2024)Krupp, Steinert, Kiefer-Emmanouilidis, Avila, Lukowicz, Kuhn, K{\"u}chemann, and Karolus}]{krupp2024challenges}
Lars Krupp, Steffen Steinert, Maximilian Kiefer-Emmanouilidis, Karina~E Avila, Paul Lukowicz, Jochen Kuhn, Stefan K{\"u}chemann, and Jakob Karolus. 2024.
\newblock Challenges and opportunities of moderating usage of large language models in education.
\newblock In \emph{Proceedings of the International Conference on Mobile and Ubiquitous Multimedia}, pages 249--254.

\bibitem[{Ku et~al.(2025)Ku, Chong, Leung, Shah, Yu, and Chen}]{ku2025theoremexplainagentmultimodalexplanationsllm}
Max Ku, Thomas Chong, Jonathan Leung, Krish Shah, Alvin Yu, and Wenhu Chen. 2025.
\newblock \href {https://arxiv.org/abs/2502.19400} {Theoremexplainagent: Towards multimodal explanations for llm theorem understanding}.
\newblock \emph{Preprint}, arXiv:2502.19400.

\bibitem[{K{\"u}chemann et~al.(2025)K{\"u}chemann, Avila, Dinc, Hortmann, Revenga, Ruf, Stausberg, Steinert, Fischer, Fischer et~al.}]{kuchemann2025opportunities}
Stefan K{\"u}chemann, Karina~E Avila, Yavuz Dinc, Chiara Hortmann, Natalia Revenga, Verena Ruf, Niklas Stausberg, Steffen Steinert, Frank Fischer, Martin Fischer, and 1 others. 2025.
\newblock On opportunities and challenges of large multimodal foundation models in education.
\newblock \emph{npj Science of Learning}, 10(1):11.

\bibitem[{Laak and Aru(2024)}]{laak2024ai}
Kristjan-Julius Laak and Jaan Aru. 2024.
\newblock Ai and personalized learning: bridging the gap with modern educational goals.
\newblock \emph{arXiv preprint arXiv:2404.02798}.

\bibitem[{Lagakis and Demetriadis(2024)}]{lagakis2024evaai}
Paraskevas Lagakis and Stavros Demetriadis. 2024.
\newblock Evaai: a multi-agent framework leveraging large language models for enhanced automated grading.
\newblock In \emph{International Conference on Intelligent Tutoring Systems}, pages 378--385. Springer.

\bibitem[{Lai et~al.(2024)Lai, Gan, Wu, Qi, and Philip}]{lai2024large}
Jinqi Lai, Wensheng Gan, Jiayang Wu, Zhenlian Qi, and S~Yu Philip. 2024.
\newblock Large language models in law: A survey.
\newblock \emph{AI Open}.

\bibitem[{L{\'a}la et~al.(2023)L{\'a}la, O'Donoghue, Shtedritski, Cox, Rodriques, and White}]{lala2023paperqa}
Jakub L{\'a}la, Odhran O'Donoghue, Aleksandar Shtedritski, Sam Cox, Samuel~G Rodriques, and Andrew~D White. 2023.
\newblock Paperqa: Retrieval-augmented generative agent for scientific research.
\newblock \emph{arXiv preprint arXiv:2312.07559}.

\bibitem[{Lee and Song(2024)}]{lee2024teachers}
Soohwan Lee and Ki-Sang Song. 2024.
\newblock Teachers' and students' perceptions of ai-generated concept explanations: Implications for integrating generative ai in computer science education.
\newblock \emph{Computers and Education: Artificial Intelligence}, 7:100283.

\bibitem[{Lei et~al.(2025)Lei, Zhang, Zuo, Payani, and Ding}]{lei2025macm}
Bin Lei, Yi~Zhang, Shan Zuo, Ali Payani, and Caiwen Ding. 2025.
\newblock Macm: Utilizing a multi-agent system for condition mining in solving complex mathematical problems.
\newblock \emph{Advances in Neural Information Processing Systems}, 37:53418--53437.

\bibitem[{Li et~al.(2024{\natexlab{a}})Li, Chen, Yang, Ai, Jia, Liu, Lin, Wu, Yuan, Hu et~al.}]{li2024legalagentbench}
Haitao Li, Junjie Chen, Jingli Yang, Qingyao Ai, Wei Jia, Youfeng Liu, Kai Lin, Yueyue Wu, Guozhi Yuan, Yiran Hu, and 1 others. 2024{\natexlab{a}}.
\newblock Legalagentbench: Evaluating llm agents in legal domain.
\newblock \emph{arXiv preprint arXiv:2412.17259}.

\bibitem[{Li et~al.(2025{\natexlab{a}})Li, Yu, Cong, Dang, Zhan, Liu, and Liu}]{li2025exploring}
Haoxuan Li, Jifan Yu, Xin Cong, Yang Dang, Yisi Zhan, Huiqin Liu, and Zhiyuan Liu. 2025{\natexlab{a}}.
\newblock Exploring llm-based student simulation for metacognitive cultivation.
\newblock \emph{arXiv preprint arXiv:2502.11678}.

\bibitem[{Li et~al.(2024{\natexlab{b}})Li, Lai, Li, Ren, Zhang, Kang, Wang, Li, Zhang, Ma et~al.}]{li2024agent}
Junkai Li, Yunghwei Lai, Weitao Li, Jingyi Ren, Meng Zhang, Xinhui Kang, Siyu Wang, Peng Li, Ya-Qin Zhang, Weizhi Ma, and 1 others. 2024{\natexlab{b}}.
\newblock Agent hospital: A simulacrum of hospital with evolvable medical agents.
\newblock \emph{arXiv preprint arXiv:2405.02957}.

\bibitem[{Li et~al.(2025{\natexlab{b}})Li, Zhao, Wang, Wang, Zhou, Srivastava, Gokmen, Lee, Li, Zhang et~al.}]{li2025embodied}
Manling Li, Shiyu Zhao, Qineng Wang, Kangrui Wang, Yu~Zhou, Sanjana Srivastava, Cem Gokmen, Tony Lee, Erran~Li Li, Ruohan Zhang, and 1 others. 2025{\natexlab{b}}.
\newblock Embodied agent interface: Benchmarking llms for embodied decision making.
\newblock \emph{Advances in Neural Information Processing Systems}, 37:100428--100534.

\bibitem[{Li et~al.(2024{\natexlab{c}})Li, Xia, Du, Zhang, Zhang, Tang, and Yu}]{li2024learning}
Qingyao Li, Wei Xia, Kounianhua Du, Qiji Zhang, Weinan Zhang, Ruiming Tang, and Yong Yu. 2024{\natexlab{c}}.
\newblock Learning structure and knowledge aware representation with large language models for concept recommendation.
\newblock \emph{arXiv preprint arXiv:2405.12442}.

\bibitem[{Li et~al.(2024{\natexlab{d}})Li, Qin, Huang, Ye, Li, Qin, Hu, Jiang, Zheng, and Yu}]{li2024rethinking}
Yinghui Li, Shang Qin, Haojing Huang, Jingheng Ye, Yangning Li, Libo Qin, Xuming Hu, Wenhao Jiang, Hai-Tao Zheng, and Philip~S Yu. 2024{\natexlab{d}}.
\newblock Rethinking the roles of large language models in chinese grammatical error correction.
\newblock \emph{arXiv preprint arXiv:2402.11420}.

\bibitem[{Li and Xu(2023)}]{li2023designing}
Zhixin Li and Ying Xu. 2023.
\newblock Designing a realistic peer-like embodied conversational agent for supporting children's storytelling.
\newblock \emph{arXiv preprint arXiv:2304.09399}.

\bibitem[{Li et~al.(2024{\natexlab{e}})Li, Zang, Ma, Guo, Zheng, Liu, Niu, Wang, Yang, Liu et~al.}]{li2024autokaggle}
Ziming Li, Qianbo Zang, David Ma, Jiawei Guo, Tuney Zheng, Minghao Liu, Xinyao Niu, Yue Wang, Jian Yang, Jiaheng Liu, and 1 others. 2024{\natexlab{e}}.
\newblock Autokaggle: A multi-agent framework for autonomous data science competitions.
\newblock \emph{arXiv preprint arXiv:2410.20424}.

\bibitem[{Liang et~al.(2023)Liang, Wang, Huang, Wu, Wu, Lu, Ma, and Li}]{liang2023unleashing}
Xinnian Liang, Bing Wang, Hui Huang, Shuangzhi Wu, Peihao Wu, Lu~Lu, Zejun Ma, and Zhoujun Li. 2023.
\newblock Unleashing infinite-length input capacity for large-scale language models with self-controlled memory system.
\newblock \emph{arXiv preprint arXiv:2304.13343}.

\bibitem[{Liang et~al.(2024)Liang, Tao, Xia, Shi, Wang, and Yang}]{liang2024cmat}
Xuechen Liang, Meiling Tao, Yinghui Xia, Tianyu Shi, Jun Wang, and JingSong Yang. 2024.
\newblock Cmat: A multi-agent collaboration tuning framework for enhancing small language models.
\newblock \emph{arXiv preprint arXiv:2404.01663}.

\bibitem[{Liem et~al.(2023)Liem, Kusnick, Beck, Windhager, and Mayr}]{liem2023workflow}
Johannes Liem, Jakob Kusnick, Samuel Beck, Florian Windhager, and Eva Mayr. 2023.
\newblock A workflow approach to visualization-based storytelling with cultural heritage data.
\newblock In \emph{2023 IEEE 8th Workshop on Visualization for the Digital Humanities (VIS4DH)}, pages 13--17. IEEE.

\bibitem[{Lim et~al.(2024)Lim, Schm{\"a}lzle, and Bente}]{lim2024artificial}
Sue Lim, Ralf Schm{\"a}lzle, and Gary Bente. 2024.
\newblock Artificial social influence via human-embodied ai agent interaction in immersive virtual reality (vr): Effects of similarity-matching during health conversations.
\newblock \emph{arXiv preprint arXiv:2406.05486}.

\bibitem[{Liu et~al.(2024{\natexlab{a}})Liu, Huang, Chen, Adarkwah, Zhang, Li, Zhang, and Da}]{liu2024personalized}
Dejian Liu, Ronghuai Huang, Ying Chen, Michael~Agyemang Adarkwah, Xiangling Zhang, Xin Li, Junjie Zhang, and Ting Da. 2024{\natexlab{a}}.
\newblock Personalized tutoring through conversational agents.
\newblock In \emph{Using Educational Robots to Enhance Learning: An Analysis of 100 Academic Articles}, pages 59--85. Springer.

\bibitem[{Liu et~al.(2024{\natexlab{b}})Liu, Zenke, Liu, Holmes, Thornton, and Malan}]{liu2024teaching}
Rongxin Liu, Carter Zenke, Charlie Liu, Andrew Holmes, Patrick Thornton, and David~J Malan. 2024{\natexlab{b}}.
\newblock Teaching cs50 with ai: leveraging generative artificial intelligence in computer science education.
\newblock In \emph{Proceedings of the 55th ACM technical symposium on computer science education V. 1}, pages 750--756.

\bibitem[{Liu et~al.(2025)Liu, Yao, Jia, Casper, Baracaldo, Hase, Yao, Liu, Xu, Li et~al.}]{liu2025rethinking}
Sijia Liu, Yuanshun Yao, Jinghan Jia, Stephen Casper, Nathalie Baracaldo, Peter Hase, Yuguang Yao, Chris~Yuhao Liu, Xiaojun Xu, Hang Li, and 1 others. 2025.
\newblock Rethinking machine unlearning for large language models.
\newblock \emph{Nature Machine Intelligence}, pages 1--14.

\bibitem[{Liu et~al.(2024{\natexlab{c}})Liu, Yin, Lin, and Chen}]{liu-etal-2024-personality}
Zhengyuan Liu, Stella~Xin Yin, Geyu Lin, and Nancy~F. Chen. 2024{\natexlab{c}}.
\newblock \href {https://doi.org/10.18653/v1/2024.emnlp-main.37} {Personality-aware student simulation for conversational intelligent tutoring systems}.
\newblock In \emph{Proceedings of the 2024 Conference on Empirical Methods in Natural Language Processing}, pages 626--642, Miami, Florida, USA. Association for Computational Linguistics.

\bibitem[{Liu et~al.(2024{\natexlab{d}})Liu, Li, Chen, Zhang, and Lee}]{liu2024understanding}
Zihan Liu, Han Li, Anfan Chen, Renwen Zhang, and Yi-Chieh Lee. 2024{\natexlab{d}}.
\newblock Understanding public perceptions of ai conversational agents: A cross-cultural analysis.
\newblock In \emph{Proceedings of the 2024 CHI Conference on Human Factors in Computing Systems}, pages 1--17.

\bibitem[{Lou et~al.(2024)Lou, Xu, Wang, Du, Kamoi, Lu, Xie, Sun, Zhang, Ahn et~al.}]{lou2024aaar}
Renze Lou, Hanzi Xu, Sijia Wang, Jiangshu Du, Ryo Kamoi, Xiaoxin Lu, Jian Xie, Yuxuan Sun, Yusen Zhang, Jihyun~Janice Ahn, and 1 others. 2024.
\newblock Aaar-1.0: Assessing ai's potential to assist research.
\newblock \emph{arXiv preprint arXiv:2410.22394}.

\bibitem[{Lu et~al.(2024)Lu, Lu, Lange, Foerster, Clune, and Ha}]{lu2024ai}
Chris Lu, Cong Lu, Robert~Tjarko Lange, Jakob Foerster, Jeff Clune, and David Ha. 2024.
\newblock The ai scientist: Towards fully automated open-ended scientific discovery.
\newblock \emph{arXiv preprint arXiv:2408.06292}.

\bibitem[{Lv et~al.(2025)Lv, Liu, Gao, Zhang, Lu, and Zhu}]{lv2025genal}
Rui Lv, Qi~Liu, Weibo Gao, Haotian Zhang, Junyu Lu, and Linbo Zhu. 2025.
\newblock Genal: Generative agent for adaptive learning.
\newblock In \emph{Proceedings of the AAAI Conference on Artificial Intelligence}, volume~39, pages 577--585.

\bibitem[{M.~Bran et~al.(2024)M.~Bran, Cox, Schilter, Baldassari, White, and Schwaller}]{m2024augmenting}
Andres M.~Bran, Sam Cox, Oliver Schilter, Carlo Baldassari, Andrew~D White, and Philippe Schwaller. 2024.
\newblock Augmenting large language models with chemistry tools.
\newblock \emph{Nature Machine Intelligence}, 6(5):525--535.

\bibitem[{Ma et~al.(2024{\natexlab{a}})Ma, Wang, Guo, Sun, Tenenbaum, Rus, Gan, and Matusik}]{ma2024llm}
Pingchuan Ma, Tsun-Hsuan Wang, Minghao Guo, Zhiqing Sun, Joshua~B Tenenbaum, Daniela Rus, Chuang Gan, and Wojciech Matusik. 2024{\natexlab{a}}.
\newblock Llm and simulation as bilevel optimizers: A new paradigm to advance physical scientific discovery.
\newblock \emph{arXiv preprint arXiv:2405.09783}.

\bibitem[{Ma et~al.(2024{\natexlab{b}})Ma, Shen, Koedinger, and Wu}]{ma2024teach}
Qianou Ma, Hua Shen, Kenneth Koedinger, and Sherry~Tongshuang Wu. 2024{\natexlab{b}}.
\newblock How to teach programming in the ai era? using llms as a teachable agent for debugging.
\newblock In \emph{International Conference on Artificial Intelligence in Education}, pages 265--279. Springer.

\bibitem[{Ma et~al.(2024{\natexlab{c}})Ma, Gou, Hao, Xu, Wang, Pan, Yang, Cao, Sun, Awadalla et~al.}]{ma2024sciagent}
Yubo Ma, Zhibin Gou, Junheng Hao, Ruochen Xu, Shuohang Wang, Liangming Pan, Yujiu Yang, Yixin Cao, Aixin Sun, Hany Awadalla, and 1 others. 2024{\natexlab{c}}.
\newblock Sciagent: Tool-augmented language models for scientific reasoning.
\newblock \emph{arXiv preprint arXiv:2402.11451}.

\bibitem[{Macina et~al.(2023)Macina, Daheim, Chowdhury, Sinha, Kapur, Gurevych, and Sachan}]{macina2023mathdial}
Jakub Macina, Nico Daheim, Sankalan~Pal Chowdhury, Tanmay Sinha, Manu Kapur, Iryna Gurevych, and Mrinmaya Sachan. 2023.
\newblock Mathdial: A dialogue tutoring dataset with rich pedagogical properties grounded in math reasoning problems.
\newblock \emph{arXiv preprint arXiv:2305.14536}.

\bibitem[{Macina et~al.(2025)Macina, Daheim, Hakimi, Kapur, Gurevych, and Sachan}]{macina2025mathtutorbench}
Jakub Macina, Nico Daheim, Ido Hakimi, Manu Kapur, Iryna Gurevych, and Mrinmaya Sachan. 2025.
\newblock Mathtutorbench: A benchmark for measuring open-ended pedagogical capabilities of llm tutors.
\newblock \emph{arXiv preprint arXiv:2502.18940}.

\bibitem[{Majumder et~al.(2024)Majumder, Surana, Agarwal, Mishra, Meena, Prakhar, Vora, Khot, Sabharwal, and Clark}]{majumder2024discoverybench}
Bodhisattwa~Prasad Majumder, Harshit Surana, Dhruv Agarwal, Bhavana~Dalvi Mishra, Abhijeetsingh Meena, Aryan Prakhar, Tirth Vora, Tushar Khot, Ashish Sabharwal, and Peter Clark. 2024.
\newblock Discoverybench: Towards data-driven discovery with large language models.
\newblock \emph{arXiv preprint arXiv:2407.01725}.

\bibitem[{Matelsky et~al.(2023)Matelsky, Parodi, Liu, Lange, and Kording}]{matelsky2023large}
Jordan~K Matelsky, Felipe Parodi, Tony Liu, Richard~D Lange, and Konrad~P Kording. 2023.
\newblock A large language model-assisted education tool to provide feedback on open-ended responses.
\newblock \emph{arXiv preprint arXiv:2308.02439}.

\bibitem[{Mehrotra et~al.(2024)Mehrotra, Jorge, Jonker, and Tielman}]{mehrotra2024integrity}
Siddharth Mehrotra, Carolina~Centeio Jorge, Catholijn~M Jonker, and Myrthe~L Tielman. 2024.
\newblock Integrity-based explanations for fostering appropriate trust in ai agents.
\newblock \emph{ACM Transactions on Interactive Intelligent Systems}, 14(1):1--36.

\bibitem[{Milano et~al.(2023)Milano, McGrane, and Leonelli}]{milano2023large}
Silvia Milano, Joshua~A McGrane, and Sabina Leonelli. 2023.
\newblock Large language models challenge the future of higher education.
\newblock \emph{Nature Machine Intelligence}, 5(4):333--334.

\bibitem[{Mithun et~al.(2025)Mithun, Noriega-Atala, Merchant, and Skidmore}]{mithun2025ai}
Paul Mithun, Enrique Noriega-Atala, Nirav Merchant, and Edwin Skidmore. 2025.
\newblock Ai-verde: A gateway for egalitarian access to large language model-based resources for educational institutions.
\newblock \emph{arXiv preprint arXiv:2502.09651}.

\bibitem[{Mitra et~al.(2024)Mitra, Khanpour, Rosset, and Awadallah}]{mitra2024orca}
Arindam Mitra, Hamed Khanpour, Corby Rosset, and Ahmed Awadallah. 2024.
\newblock Orca-math: Unlocking the potential of slms in grade school math.
\newblock \emph{arXiv preprint arXiv:2402.14830}.

\bibitem[{Mo and Hu(2024)}]{mo-hu-2024-expertease}
Kaijie Mo and Renfen Hu. 2024.
\newblock \href {https://doi.org/10.18653/v1/2024.findings-emnlp.530} {{E}xpert{E}ase: A multi-agent framework for grade-specific document simplification with large language models}.
\newblock In \emph{Findings of the Association for Computational Linguistics: EMNLP 2024}, pages 9080--9099, Miami, Florida, USA. Association for Computational Linguistics.

\bibitem[{Moon et~al.(2024)Moon, Lee, Eo, Park, Seo, and Lim}]{moon2024generative}
Hyeonseok Moon, Jaewook Lee, Sugyeong Eo, Chanjun Park, Jaehyung Seo, and Heui-Seok Lim. 2024.
\newblock Generative interpretation: Toward human-like evaluation for educational question-answer pair generation.
\newblock In \emph{Findings of the Association for Computational Linguistics: EACL 2024}, pages 2185--2196.

\bibitem[{Moret et~al.(2023)Moret, Pachon~Angona, Cotos, Yan, Atz, Brunner, Baumgartner, Grisoni, and Schneider}]{moret2023leveraging}
Michael Moret, Irene Pachon~Angona, Leandro Cotos, Shen Yan, Kenneth Atz, Cyrill Brunner, Martin Baumgartner, Francesca Grisoni, and Gisbert Schneider. 2023.
\newblock Leveraging molecular structure and bioactivity with chemical language models for de novo drug design.
\newblock \emph{Nature Communications}, 14(1):114.

\bibitem[{Mower and Bou-Ammar(2025)}]{mower2025khwarizmi}
Christopher~E Mower and Haitham Bou-Ammar. 2025.
\newblock Al-khwarizmi: Discovering physical laws with foundation models.
\newblock \emph{arXiv preprint arXiv:2502.01702}.

\bibitem[{Nair et~al.(2024)Nair, Tan, Su, Gere, Wang, and Wang}]{nair2024closing}
Inderjeet Nair, Jiaye Tan, Xiaotian Su, Anne Gere, Xu~Wang, and Lu~Wang. 2024.
\newblock Closing the loop: Learning to generate writing feedback via language model simulated student revisions.
\newblock In \emph{Proceedings of the 2024 Conference on Empirical Methods in Natural Language Processing}, pages 16636--16657.

\bibitem[{Narayanan et~al.(2024)Narayanan, Braza, Griffiths, Ponnapati, Bou, Laurent, Kabeli, Wellawatte, Cox, Rodriques et~al.}]{narayanan2024aviary}
Siddharth Narayanan, James~D Braza, Ryan-Rhys Griffiths, Manu Ponnapati, Albert Bou, Jon Laurent, Ori Kabeli, Geemi Wellawatte, Sam Cox, Samuel~G Rodriques, and 1 others. 2024.
\newblock Aviary: training language agents on challenging scientific tasks.
\newblock \emph{arXiv preprint arXiv:2412.21154}.

\bibitem[{Nathani et~al.(2025)Nathani, Madaan, Roberts, Bashlykov, Menon, Moens, Budhiraja, Magka, Vorotilov, Chaurasia et~al.}]{nathani2025mlgym}
Deepak Nathani, Lovish Madaan, Nicholas Roberts, Nikolay Bashlykov, Ajay Menon, Vincent Moens, Amar Budhiraja, Despoina Magka, Vladislav Vorotilov, Gaurav Chaurasia, and 1 others. 2025.
\newblock Mlgym: A new framework and benchmark for advancing ai research agents.
\newblock \emph{arXiv preprint arXiv:2502.14499}.

\bibitem[{Nelson(2024)}]{nelson2024other}
Jack Nelson. 2024.
\newblock The other'llm': Large language models and the future of legal education.
\newblock \emph{European Journal of Legal Education}, 5(1):127--155.

\bibitem[{Ng et~al.(2024)Ng, Tan, and Leung}]{ng2024empowering}
Davy Tsz~Kit Ng, Chee~Wei Tan, and Jac Ka~Lok Leung. 2024.
\newblock Empowering student self-regulated learning and science education through chatgpt: A pioneering pilot study.
\newblock \emph{British Journal of Educational Technology}, 55(4):1328--1353.

\bibitem[{Nguyen et~al.(2022)Nguyen, Wu, Tuan, Hai, and Bing}]{nguyen2022adaptive}
Thong Nguyen, Xiaobao Wu, Luu~Anh Tuan, Zhen Hai, and Lidong Bing. 2022.
\newblock Adaptive contrastive learning on multimodal transformer for review helpfulness prediction.
\newblock In \emph{Proceedings of the 2022 Conference on Empirical Methods in Natural Language Processing}, pages 10085--10096.

\bibitem[{Pan et~al.(2024)Pan, Kitson, Wan, and Prpa}]{pan2024ellma}
Mengxu Pan, Alexandra Kitson, Hongyu Wan, and Mirjana Prpa. 2024.
\newblock Ellma-t: an embodied llm-agent for supporting english language learning in social vr.
\newblock \emph{arXiv preprint arXiv:2410.02406}.

\bibitem[{Pang et~al.(2024)Pang, Hong, Zhou, Lv, Yang, Liang, Han, and Zhang}]{pang2024physics}
Xinyu Pang, Ruixin Hong, Zhanke Zhou, Fangrui Lv, Xinwei Yang, Zhilong Liang, Bo~Han, and Changshui Zhang. 2024.
\newblock Physics reasoner: Knowledge-augmented reasoning for solving physics problems with large language models.
\newblock \emph{arXiv preprint arXiv:2412.13791}.

\bibitem[{Pardos et~al.(2023)Pardos, Tang, Anastasopoulos, Sheel, and Zhang}]{10.1145/3544548.3581574}
Zachary~A. Pardos, Matthew Tang, Ioannis Anastasopoulos, Shreya~K. Sheel, and Ethan Zhang. 2023.
\newblock \href {https://doi.org/10.1145/3544548.3581574} {Oatutor: An open-source adaptive tutoring system and curated content library for learning sciences research}.
\newblock In \emph{Proceedings of the 2023 CHI Conference on Human Factors in Computing Systems}, CHI '23, New York, NY, USA. Association for Computing Machinery.

\bibitem[{Park et~al.(2024)Park, Kim, Lee, Kwon, and Kim}]{park2024empowering}
Minju Park, Sojung Kim, Seunghyun Lee, Soonwoo Kwon, and Kyuseok Kim. 2024.
\newblock Empowering personalized learning through a conversation-based tutoring system with student modeling.
\newblock In \emph{Extended Abstracts of the CHI Conference on Human Factors in Computing Systems}, pages 1--10.

\bibitem[{Pedro et~al.(2019)Pedro, Subosa, Rivas, and Valverde}]{pedro2019artificial}
Francesc Pedro, Miguel Subosa, Axel Rivas, and Paula Valverde. 2019.
\newblock Artificial intelligence in education: Challenges and opportunities for sustainable development.

\bibitem[{Qin et~al.(2023)Qin, Liang, Ye, Zhu, Yan, Lu, Lin, Cong, Tang, Qian et~al.}]{qin2023toolllm}
Yujia Qin, Shihao Liang, Yining Ye, Kunlun Zhu, Lan Yan, Yaxi Lu, Yankai Lin, Xin Cong, Xiangru Tang, Bill Qian, and 1 others. 2023.
\newblock Toolllm: Facilitating large language models to master 16000+ real-world apis.
\newblock \emph{arXiv preprint arXiv:2307.16789}.

\bibitem[{Raihan et~al.(2025)Raihan, Siddiq, Santos, and Zampieri}]{raihan2025large}
Nishat Raihan, Mohammed~Latif Siddiq, Joanna~CS Santos, and Marcos Zampieri. 2025.
\newblock Large language models in computer science education: A systematic literature review.
\newblock In \emph{Proceedings of the 56th ACM Technical Symposium on Computer Science Education V. 1}, pages 938--944.

\bibitem[{Ramos et~al.(2025)Ramos, Collison, and White}]{ramos2025review}
Mayk~Caldas Ramos, Christopher~J Collison, and Andrew~D White. 2025.
\newblock A review of large language models and autonomous agents in chemistry.
\newblock \emph{Chemical Science}.

\bibitem[{Razafinirina et~al.(2024)Razafinirina, Dimbisoa, and Mahatody}]{razafinirina2024pedagogical}
Mahefa~Abel Razafinirina, William~Germain Dimbisoa, and Thomas Mahatody. 2024.
\newblock Pedagogical alignment of large language models (llm) for personalized learning: a survey, trends and challenges.
\newblock \emph{Journal of Intelligent Learning Systems and Applications}, 16(4):448--480.

\bibitem[{Renze and Guven(2024)}]{renze2024self}
Matthew Renze and Erhan Guven. 2024.
\newblock Self-reflection in llm agents: Effects on problem-solving performance.
\newblock \emph{arXiv preprint arXiv:2405.06682}.

\bibitem[{Roccas et~al.(2002)Roccas, Sagiv, Schwartz, and Knafo}]{roccas2002big}
Sonia Roccas, Lilach Sagiv, Shalom~H Schwartz, and Ariel Knafo. 2002.
\newblock The big five personality factors and personal values.
\newblock \emph{Personality and social psychology bulletin}, 28(6):789--801.

\bibitem[{Ruan et~al.(2024)Ruan, Lu, Xu, Zhang, Xuan, Pan, Fang, Gao, Shen, Ye et~al.}]{ruan2024accelerated}
Yixiang Ruan, Chenyin Lu, Ning Xu, Jian Zhang, Jun Xuan, Jianzhang Pan, Qun Fang, Hanyu Gao, Xiaodong Shen, Ning Ye, and 1 others. 2024.
\newblock Accelerated end-to-end chemical synthesis development with large language models.

\bibitem[{Ryan et~al.(2023)Ryan, Naous, and Xu}]{ryan-etal-2023-revisiting}
Michael~J Ryan, Tarek Naous, and Wei Xu. 2023.
\newblock \href {https://doi.org/10.18653/v1/2023.acl-long.269} {Revisiting non-{E}nglish text simplification: A unified multilingual benchmark}.
\newblock In \emph{Proceedings of the 61st Annual Meeting of the Association for Computational Linguistics (Volume 1: Long Papers)}, pages 4898--4927, Toronto, Canada. Association for Computational Linguistics.

\bibitem[{Scarlatos et~al.(2025)Scarlatos, Baker, and Lan}]{scarlatos2025exploring}
Alexander Scarlatos, Ryan~S Baker, and Andrew Lan. 2025.
\newblock Exploring knowledge tracing in tutor-student dialogues using llms.
\newblock In \emph{Proceedings of the 15th International Learning Analytics and Knowledge Conference}, pages 249--259.

\bibitem[{Schmidgall et~al.(2025)Schmidgall, Su, Wang, Sun, Wu, Yu, Liu, Liu, and Barsoum}]{schmidgall2025agent}
Samuel Schmidgall, Yusheng Su, Ze~Wang, Ximeng Sun, Jialian Wu, Xiaodong Yu, Jiang Liu, Zicheng Liu, and Emad Barsoum. 2025.
\newblock Agent laboratory: Using llm agents as research assistants.
\newblock \emph{arXiv preprint arXiv:2501.04227}.

\bibitem[{Shafiq et~al.(2022)Shafiq, Marjani, Habeeb, and Asirvatham}]{shafiq2022student}
Dalia~Abdulkareem Shafiq, Mohsen Marjani, Riyaz Ahamed~Ariyaluran Habeeb, and David Asirvatham. 2022.
\newblock Student retention using educational data mining and predictive analytics: a systematic literature review.
\newblock \emph{IEEE Access}, 10:72480--72503.

\bibitem[{Shahzad et~al.(2025)Shahzad, Mazhar, Tariq, Ahmad, Ouahada, and Hamam}]{shahzad2025comprehensive}
Tariq Shahzad, Tehseen Mazhar, Muhammad~Usman Tariq, Wasim Ahmad, Khmaies Ouahada, and Habib Hamam. 2025.
\newblock A comprehensive review of large language models: issues and solutions in learning environments.
\newblock \emph{Discover Sustainability}, 6(1):27.

\bibitem[{Shen et~al.(2024)Shen, Chen, Mamalakis, Liu, Li, Su, He, Li{\`o}, and Wang}]{shen2024toursynbio}
Yiqing Shen, Zan Chen, Michail Mamalakis, Yungeng Liu, Tianbin Li, Yanzhou Su, Junjun He, Pietro Li{\`o}, and Yu~Guang Wang. 2024.
\newblock Toursynbio: A multi-modal large model and agent framework to bridge text and protein sequences for protein engineering.
\newblock In \emph{2024 IEEE International Conference on Bioinformatics and Biomedicine (BIBM)}, pages 2382--2389. IEEE.

\bibitem[{Shojaee et~al.(2024)Shojaee, Meidani, Gupta, Farimani, and Reddy}]{shojaee2024llm}
Parshin Shojaee, Kazem Meidani, Shashank Gupta, Amir~Barati Farimani, and Chandan~K Reddy. 2024.
\newblock Llm-sr: Scientific equation discovery via programming with large language models.
\newblock \emph{arXiv preprint arXiv:2404.18400}.

\bibitem[{Shusterman et~al.(2025)Shusterman, Waters, O’Neill, Bangs, Luu, and Tucker}]{shusterman2025active}
Roma Shusterman, Allison~C Waters, Shannon O’Neill, Marshall Bangs, Phan Luu, and Don~M Tucker. 2025.
\newblock An active inference strategy for prompting reliable responses from large language models in medical practice.
\newblock \emph{npj Digital Medicine}, 8(1):119.

\bibitem[{Si et~al.(2023)Si, Ma, Gao, Wu, Lin, Dai, Li, Yan, Huang, and Li}]{si2023spokenwoz}
Shuzheng Si, Wentao Ma, Haoyu Gao, Yuchuan Wu, Ting-En Lin, Yinpei Dai, Hangyu Li, Rui Yan, Fei Huang, and Yongbin Li. 2023.
\newblock Spokenwoz: A large-scale speech-text benchmark for spoken task-oriented dialogue agents.
\newblock \emph{Advances in Neural Information Processing Systems}, 36:39088--39118.

\bibitem[{Simon and Muise(2022)}]{simon2022tattletale}
Nisha Simon and Christian Muise. 2022.
\newblock Tattletale: storytelling with planning and large language models.
\newblock In \emph{ICAPS Workshop on Scheduling and Planning Applications}.

\bibitem[{Skarlinski et~al.(2024)Skarlinski, Cox, Laurent, Braza, Hinks, Hammerling, Ponnapati, Rodriques, and White}]{skarlinski2024language}
Michael~D Skarlinski, Sam Cox, Jon~M Laurent, James~D Braza, Michaela Hinks, Michael~J Hammerling, Manvitha Ponnapati, Samuel~G Rodriques, and Andrew~D White. 2024.
\newblock Language agents achieve superhuman synthesis of scientific knowledge.
\newblock \emph{arXiv preprint arXiv:2409.13740}.

\bibitem[{{\v{S}}kobo and Petri{\v{c}}evi{\'c}(2023)}]{vskobo2023navigating}
Milena {\v{S}}kobo and Vedran Petri{\v{c}}evi{\'c}. 2023.
\newblock Navigating the challenges and opportunities of literary translation in the age of ai: Striking a balance between human expertise and machine power.
\newblock \emph{Dru{\v{s}}tvene i humanisti{\v{c}}ke studije}, 8(2 (23)):317--336.

\bibitem[{Sohn et~al.(2024)Sohn, Li, Zhang, Chang, and Kapadia}]{sohn2024words}
Samuel~S Sohn, Danrui Li, Sen Zhang, Che-Jui Chang, and Mubbasir Kapadia. 2024.
\newblock From words to worlds: Transforming one-line prompt into immersive multi-modal digital stories with communicative llm agent.
\newblock \emph{arXiv preprint arXiv:2406.10478}.

\bibitem[{Song et~al.(2024)Song, Luo, Chen, Huang, Zhu, Liu, Zhang, Zou, Zhang, Shang et~al.}]{song2024multi}
Tao Song, Man Luo, Linjiang Chen, Yan Huang, Qing Zhu, Daobin Liu, Baicheng Zhang, Gang Zou, Fei Zhang, Weiwei Shang, and 1 others. 2024.
\newblock A multi-agent-driven robotic ai chemist enabling autonomous chemical research on demand.

\bibitem[{Song et~al.(2022)Song, Li, Cai, Yang, Yang, and Liu}]{song2022survey}
Xiangyu Song, Jianxin Li, Taotao Cai, Shuiqiao Yang, Tingting Yang, and Chengfei Liu. 2022.
\newblock A survey on deep learning based knowledge tracing.
\newblock \emph{Knowledge-Based Systems}, 258:110036.

\bibitem[{Sonlu et~al.(2024)Sonlu, Bendiksen, Durupinar, and G{\"u}d{\"u}kbay}]{sonlu2024effects}
Sinan Sonlu, Bennie Bendiksen, Funda Durupinar, and U{\u{g}}ur G{\"u}d{\"u}kbay. 2024.
\newblock The effects of embodiment and personality expression on learning in llm-based educational agents.
\newblock \emph{arXiv preprint arXiv:2407.10993}.

\bibitem[{Sprueill et~al.(2024)Sprueill, Edwards, Agarwal, Olarte, Sanyal, Johnston, Liu, Ji, and Choudhury}]{sprueill2024chemreasoner}
Henry~W Sprueill, Carl Edwards, Khushbu Agarwal, Mariefel~V Olarte, Udishnu Sanyal, Conrad Johnston, Hongbin Liu, Heng Ji, and Sutanay Choudhury. 2024.
\newblock Chemreasoner: Heuristic search over a large language model's knowledge space using quantum-chemical feedback.
\newblock \emph{arXiv preprint arXiv:2402.10980}.

\bibitem[{Sripathi et~al.(2024)Sripathi, Moscarella, Steele, Yoho, You, Prevost, Urban-Lurain, Merrill, and Haudek}]{sripathi2024machine}
Kamali~N Sripathi, Rosa~A Moscarella, Matthew Steele, Rachel Yoho, Hyesun You, Luanna~B Prevost, Mark Urban-Lurain, John Merrill, and Kevin~C Haudek. 2024.
\newblock Machine learning mixed methods text analysis: An illustration from automated scoring models of student writing in biology education.
\newblock \emph{Journal of mixed methods research}, 18(1):48--70.

\bibitem[{Steenstra et~al.(2024)Steenstra, Murali, Perkins, Joseph, Paasche-Orlow, and Bickmore}]{steenstra2024engaging}
Ian Steenstra, Prasanth Murali, Rebecca~B Perkins, Natalie Joseph, Michael~K Paasche-Orlow, and Timothy Bickmore. 2024.
\newblock Engaging and entertaining adolescents in health education using llm-generated fantasy narrative games and virtual agents.
\newblock In \emph{Extended Abstracts of the CHI Conference on Human Factors in Computing Systems}, pages 1--8.

\bibitem[{Su et~al.(2025)Su, Yan, Fu, Zhang, Ye, Liu, Huo, Zhou, and Hu}]{su2025essayjudge}
Jiamin Su, Yibo Yan, Fangteng Fu, Han Zhang, Jingheng Ye, Xiang Liu, Jiahao Huo, Huiyu Zhou, and Xuming Hu. 2025.
\newblock Essayjudge: A multi-granular benchmark for assessing automated essay scoring capabilities of multimodal large language models.
\newblock \emph{arXiv preprint arXiv:2502.11916}.

\bibitem[{Sumers et~al.()Sumers, Yao, Narasimhan, and Griffiths}]{sumerscognitive}
Theodore Sumers, Shunyu Yao, Karthik Narasimhan, and Thomas Griffiths.
\newblock Cognitive architectures for language agents.
\newblock \emph{Transactions on Machine Learning Research}.

\bibitem[{Sun et~al.(2024)Sun, Dai, Luo, Chang, and Li}]{sun2024lawluo}
Jingyun Sun, Chengxiao Dai, Zhongze Luo, Yangbo Chang, and Yang Li. 2024.
\newblock Lawluo: A chinese law firm co-run by llm agents.
\newblock \emph{arXiv preprint arXiv:2407.16252}.

\bibitem[{Swan et~al.(2023)Swan, Kido, Roland, and Santos}]{swan2023math}
Melanie Swan, Takashi Kido, Eric Roland, and Renato P~dos Santos. 2023.
\newblock Math agents: Computational infrastructure, mathematical embedding, and genomics.
\newblock \emph{arXiv preprint arXiv:2307.02502}.

\bibitem[{Tan et~al.(2023)Tan, Pang, Fan, and Yu}]{tan2023towards}
Kehui Tan, Tianqi Pang, Chenyou Fan, and Song Yu. 2023.
\newblock Towards applying powerful large ai models in classroom teaching: Opportunities, challenges and prospects.
\newblock \emph{arXiv preprint arXiv:2305.03433}.

\bibitem[{Tang et~al.(2023)Tang, Liu, Cai, Shao, Lu, Zhang, Deng, Hu, An, Huang et~al.}]{tang2023ml}
Xiangru Tang, Yuliang Liu, Zefan Cai, Yanjun Shao, Junjie Lu, Yichi Zhang, Zexuan Deng, Helan Hu, Kaikai An, Ruijun Huang, and 1 others. 2023.
\newblock Ml-bench: Evaluating large language models and agents for machine learning tasks on repository-level code.
\newblock \emph{arXiv preprint arXiv:2311.09835}.

\bibitem[{Tian et~al.(2024{\natexlab{a}})Tian, Gao, Zhang, Chen, Fan, Guo, Haas, Ji, Krongchon, Li et~al.}]{tian2024scicode}
Minyang Tian, Luyu Gao, Shizhuo Zhang, Xinan Chen, Cunwei Fan, Xuefei Guo, Roland Haas, Pan Ji, Kittithat Krongchon, Yao Li, and 1 others. 2024{\natexlab{a}}.
\newblock Scicode: A research coding benchmark curated by scientists.
\newblock \emph{Advances in Neural Information Processing Systems}, 37:30624--30650.

\bibitem[{Tian et~al.(2024{\natexlab{b}})Tian, Jin, Yeganova, Lai, Zhu, Chen, Yang, Chen, Kim, Comeau et~al.}]{tian2024opportunities}
Shubo Tian, Qiao Jin, Lana Yeganova, Po-Ting Lai, Qingqing Zhu, Xiuying Chen, Yifan Yang, Qingyu Chen, Won Kim, Donald~C Comeau, and 1 others. 2024{\natexlab{b}}.
\newblock Opportunities and challenges for chatgpt and large language models in biomedicine and health.
\newblock \emph{Briefings in Bioinformatics}, 25(1):bbad493.

\bibitem[{Tsai et~al.(2023)Tsai, Ong, and Chen}]{tsai2023exploring}
Meng-Lin Tsai, Chong~Wei Ong, and Cheng-Liang Chen. 2023.
\newblock Exploring the use of large language models (llms) in chemical engineering education: Building core course problem models with chat-gpt.
\newblock \emph{Education for Chemical Engineers}, 44:71--95.

\bibitem[{Ullah et~al.(2024)Ullah, Parwani, Baig, and Singh}]{ullah2024challenges}
Ehsan Ullah, Anil Parwani, Mirza~Mansoor Baig, and Rajendra Singh. 2024.
\newblock Challenges and barriers of using large language models (llm) such as chatgpt for diagnostic medicine with a focus on digital pathology--a recent scoping review.
\newblock \emph{Diagnostic pathology}, 19(1):43.

\bibitem[{Valmeekam et~al.(2023)Valmeekam, Marquez, Olmo, Sreedharan, and Kambhampati}]{valmeekam2023planbench}
Karthik Valmeekam, Matthew Marquez, Alberto Olmo, Sarath Sreedharan, and Subbarao Kambhampati. 2023.
\newblock Planbench: An extensible benchmark for evaluating large language models on planning and reasoning about change.
\newblock \emph{Advances in Neural Information Processing Systems}, 36:38975--38987.

\bibitem[{Voultsiou and Moussiades(2025)}]{voultsiou2025systematic}
Evdokia Voultsiou and Lefteris Moussiades. 2025.
\newblock A systematic review of ai, vr, and llm applications in special education: Opportunities, challenges, and future directions.
\newblock \emph{Education and Information Technologies}, pages 1--41.

\bibitem[{Wan et~al.(2024)Wan, Wu, Chen, and Li}]{wan2024cot}
Guangya Wan, Yuqi Wu, Jie Chen, and Sheng Li. 2024.
\newblock Cot rerailer: Enhancing the reliability of large language models in complex reasoning tasks through error detection and correction.
\newblock \emph{arXiv preprint arXiv:2408.13940}.

\bibitem[{Wang et~al.(2022)Wang, Liu, Chen, Huang, Yin, Wang, and Su}]{wang2022neuralcd}
Fei Wang, Qi~Liu, Enhong Chen, Zhenya Huang, Yu~Yin, Shijin Wang, and Yu~Su. 2022.
\newblock Neuralcd: a general framework for cognitive diagnosis.
\newblock \emph{IEEE Transactions on Knowledge and Data Engineering}, 35(8):8312--8327.

\bibitem[{Wang et~al.(2025{\natexlab{a}})Wang, Dai, Zhang, Ma, Li, and Chai}]{wang2025training}
Jian Wang, Yinpei Dai, Yichi Zhang, Ziqiao Ma, Wenjie Li, and Joyce Chai. 2025{\natexlab{a}}.
\newblock Training turn-by-turn verifiers for dialogue tutoring agents: The curious case of llms as your coding tutors.
\newblock \emph{arXiv preprint arXiv:2502.13311}.

\bibitem[{Wang et~al.(2024{\natexlab{a}})Wang, Ma, Feng, Zhang, Yang, Zhang, Chen, Tang, Chen, Lin et~al.}]{wang2024survey}
Lei Wang, Chen Ma, Xueyang Feng, Zeyu Zhang, Hao Yang, Jingsen Zhang, Zhiyuan Chen, Jiakai Tang, Xu~Chen, Yankai Lin, and 1 others. 2024{\natexlab{a}}.
\newblock A survey on large language model based autonomous agents.
\newblock \emph{Frontiers of Computer Science}, 18(6):186345.

\bibitem[{Wang et~al.(2024{\natexlab{b}})Wang, Du, Jiao, Lyu, Pang, Cui, Song, Wong, Shi, and Tu}]{wang-etal-2024-benchmarking}
Longyue Wang, Zefeng Du, Wenxiang Jiao, Chenyang Lyu, Jianhui Pang, Leyang Cui, Kaiqiang Song, Derek Wong, Shuming Shi, and Zhaopeng Tu. 2024{\natexlab{b}}.
\newblock \href {https://doi.org/10.18653/v1/2024.findings-acl.428} {Benchmarking and improving long-text translation with large language models}.
\newblock In \emph{Findings of the Association for Computational Linguistics: ACL 2024}, pages 7175--7187, Bangkok, Thailand. Association for Computational Linguistics.

\bibitem[{Wang et~al.(2025{\natexlab{b}})Wang, Zhan, Lian, Hu, Yuan, Zhang, Xie, and Xiong}]{wang2025llm}
Tianfu Wang, Yi~Zhan, Jianxun Lian, Zhengyu Hu, Nicholas~Jing Yuan, Qi~Zhang, Xing Xie, and Hui Xiong. 2025{\natexlab{b}}.
\newblock Llm-powered multi-agent framework for goal-oriented learning in intelligent tutoring system.
\newblock \emph{arXiv preprint arXiv:2501.15749}.

\bibitem[{Wang et~al.(2024{\natexlab{c}})Wang, Chen, Jia, Wang, Fang, Wang, Gao, Xie, Xu, Dai et~al.}]{wang2024weaver}
Tiannan Wang, Jiamin Chen, Qingrui Jia, Shuai Wang, Ruoyu Fang, Huilin Wang, Zhaowei Gao, Chunzhao Xie, Chuou Xu, Jihong Dai, and 1 others. 2024{\natexlab{c}}.
\newblock Weaver: Foundation models for creative writing.
\newblock \emph{arXiv preprint arXiv:2401.17268}.

\bibitem[{Wang et~al.(2024{\natexlab{d}})Wang, Chen, Dingjie, Zhiyi, Chen, Xiao, Chen, Jiang, Li, Wan, Wang, and Li}]{wang-etal-2024-cmb}
Xidong Wang, Guiming Chen, Song Dingjie, Zhang Zhiyi, Zhihong Chen, Qingying Xiao, Junying Chen, Feng Jiang, Jianquan Li, Xiang Wan, Benyou Wang, and Haizhou Li. 2024{\natexlab{d}}.
\newblock \href {https://doi.org/10.18653/v1/2024.naacl-long.343} {{CMB}: A comprehensive medical benchmark in {C}hinese}.
\newblock In \emph{Proceedings of the 2024 Conference of the North American Chapter of the Association for Computational Linguistics: Human Language Technologies (Volume 1: Long Papers)}, pages 6184--6205, Mexico City, Mexico. Association for Computational Linguistics.

\bibitem[{Wang et~al.(2024{\natexlab{e}})Wang, Wang, Tsutsui, Lin, Wen, and Kot}]{wang2024evolving}
Xiyu Wang, Yufei Wang, Satoshi Tsutsui, Weisi Lin, Bihan Wen, and Alex Kot. 2024{\natexlab{e}}.
\newblock Evolving storytelling: benchmarks and methods for new character customization with diffusion models.
\newblock In \emph{Proceedings of the 32nd ACM International Conference on Multimedia}, pages 3751--3760.

\bibitem[{Wang et~al.(2023)Wang, Duan, Fox, and Srinivasa}]{wang2023newton}
Yi~Ru Wang, Jiafei Duan, Dieter Fox, and Siddhartha Srinivasa. 2023.
\newblock Newton: Are large language models capable of physical reasoning?
\newblock \emph{arXiv preprint arXiv:2310.07018}.

\bibitem[{Wei et~al.(2024)Wei, Qiu, Yu, and Yuan}]{wei2024medco}
Hao Wei, Jianing Qiu, Haibao Yu, and Wu~Yuan. 2024.
\newblock Medco: Medical education copilots based on a multi-agent framework.
\newblock \emph{arXiv preprint arXiv:2408.12496}.

\bibitem[{Weng(2023)}]{weng2023prompt}
Lilian Weng. 2023.
\newblock Llm-powered autonomous agents.
\newblock \emph{lilianweng.github.io}.

\bibitem[{Wu et~al.(2024{\natexlab{a}})Wu, Yuan, Haffari, and Wang}]{wu2024perhaps}
Minghao Wu, Yulin Yuan, Gholamreza Haffari, and Longyue Wang. 2024{\natexlab{a}}.
\newblock (perhaps) beyond human translation: Harnessing multi-agent collaboration for translating ultra-long literary texts.
\newblock \emph{arXiv preprint arXiv:2405.11804}.

\bibitem[{Wu et~al.(2023{\natexlab{a}})Wu, Bansal, Zhang, Wu, Li, Zhu, Jiang, Zhang, Zhang, Liu et~al.}]{wu2023autogen}
Qingyun Wu, Gagan Bansal, Jieyu Zhang, Yiran Wu, Beibin Li, Erkang Zhu, Li~Jiang, Xiaoyun Zhang, Shaokun Zhang, Jiale Liu, and 1 others. 2023{\natexlab{a}}.
\newblock Autogen: Enabling next-gen llm applications via multi-agent conversation.
\newblock \emph{arXiv preprint arXiv:2308.08155}.

\bibitem[{Wu et~al.(2024{\natexlab{b}})Wu, Zhao, Zhu, Shi, Yang, Liu, Zhai, Yao, Li, Du et~al.}]{wu2024usable}
Xuansheng Wu, Haiyan Zhao, Yaochen Zhu, Yucheng Shi, Fan Yang, Tianming Liu, Xiaoming Zhai, Wenlin Yao, Jundong Li, Mengnan Du, and 1 others. 2024{\natexlab{b}}.
\newblock Usable xai: 10 strategies towards exploiting explainability in the llm era.
\newblock \emph{arXiv preprint arXiv:2403.08946}.

\bibitem[{Wu(2025)}]{wu2025empirical}
Yiran Wu. 2025.
\newblock An empirical study on challenging math problem solving with llm-based conversational agents.

\bibitem[{Wu et~al.(2023{\natexlab{b}})Wu, Jia, Zhang, Li, Zhu, Wang, Lee, Peng, Wu, and Wang}]{wu2023mathchat}
Yiran Wu, Feiran Jia, Shaokun Zhang, Hangyu Li, Erkang Zhu, Yue Wang, Yin~Tat Lee, Richard Peng, Qingyun Wu, and Chi Wang. 2023{\natexlab{b}}.
\newblock Mathchat: Converse to tackle challenging math problems with llm agents.
\newblock \emph{arXiv preprint arXiv:2306.01337}.

\bibitem[{Xia et~al.(2024)Xia, Deng, Dunn, and Zhang}]{xia2024agentless}
Chunqiu~Steven Xia, Yinlin Deng, Soren Dunn, and Lingming Zhang. 2024.
\newblock Agentless: Demystifying llm-based software engineering agents.
\newblock \emph{arXiv preprint arXiv:2407.01489}.

\bibitem[{Xie et~al.(2024{\natexlab{a}})Xie, Bai, Gao, Xue, Fang, Zhao, Li, Zhu, Ni, and Yang}]{xie2024delilaw}
Nan Xie, Yuelin Bai, Hengyuan Gao, Ziqiang Xue, Feiteng Fang, Qixuan Zhao, Zhijian Li, Liang Zhu, Shiwen Ni, and Min Yang. 2024{\natexlab{a}}.
\newblock Delilaw: A chinese legal counselling system based on a large language model.
\newblock In \emph{Proceedings of the 33rd ACM International Conference on Information and Knowledge Management}, pages 5299--5303.

\bibitem[{Xie et~al.(2024{\natexlab{b}})Xie, Zhang, Chen, Li, Zhao, Cao, Toh, Cheng, Shin, Lei et~al.}]{xie2024osworld}
Tianbao Xie, Danyang Zhang, Jixuan Chen, Xiaochuan Li, Siheng Zhao, Ruisheng Cao, Jing~Hua Toh, Zhoujun Cheng, Dongchan Shin, Fangyu Lei, and 1 others. 2024{\natexlab{b}}.
\newblock Osworld: Benchmarking multimodal agents for open-ended tasks in real computer environments.
\newblock \emph{Advances in Neural Information Processing Systems}, 37:52040--52094.

\bibitem[{Xiong et~al.(2024)Xiong, Shi, Shen, Rosenberg, Qin, Calandriello, Khalman, Joshi, Piot, Saleh et~al.}]{xiong2024building}
Wei Xiong, Chengshuai Shi, Jiaming Shen, Aviv Rosenberg, Zhen Qin, Daniele Calandriello, Misha Khalman, Rishabh Joshi, Bilal Piot, Mohammad Saleh, and 1 others. 2024.
\newblock Building math agents with multi-turn iterative preference learning.
\newblock \emph{arXiv preprint arXiv:2409.02392}.

\bibitem[{Xu et~al.(2025)Xu, Wen, Pan, Dominguez, Hu, and Zhang}]{xu2025classroom}
Songlin Xu, Hao-Ning Wen, Hongyi Pan, Dallas Dominguez, Dongyin Hu, and Xinyu Zhang. 2025.
\newblock Classroom simulacra: Building contextual student generative agents in online education for learning behavioral simulation.
\newblock \emph{arXiv preprint arXiv:2502.02780}.

\bibitem[{Xu et~al.(2024{\natexlab{a}})Xu, Zhang, and Qin}]{xu2024eduagent}
Songlin Xu, Xinyu Zhang, and Lianhui Qin. 2024{\natexlab{a}}.
\newblock Eduagent: Generative student agents in learning.
\newblock \emph{arXiv preprint arXiv:2404.07963}.

\bibitem[{Xu et~al.(2024{\natexlab{b}})Xu, Zhang, Chu, Wang, and Wen}]{xu2024ai}
Tianlong Xu, Yi-Fan Zhang, Zhendong Chu, Shen Wang, and Qingsong Wen. 2024{\natexlab{b}}.
\newblock Ai-driven virtual teacher for enhanced educational efficiency: Leveraging large pretrain models for autonomous error analysis and correction.
\newblock \emph{arXiv preprint arXiv:2409.09403}.

\bibitem[{Yan et~al.(2024{\natexlab{a}})Yan, Sha, Zhao, Li, Martinez-Maldonado, Chen, Li, Jin, and Ga{\v{s}}evi{\'c}}]{yan2024practical}
Lixiang Yan, Lele Sha, Linxuan Zhao, Yuheng Li, Roberto Martinez-Maldonado, Guanliang Chen, Xinyu Li, Yueqiao Jin, and Dragan Ga{\v{s}}evi{\'c}. 2024{\natexlab{a}}.
\newblock Practical and ethical challenges of large language models in education: A systematic scoping review.
\newblock \emph{British Journal of Educational Technology}, 55(1):90--112.

\bibitem[{Yan and Lee(2024)}]{yan2024georeasoner}
Yibo Yan and Joey Lee. 2024.
\newblock Georeasoner: Reasoning on geospatially grounded context for natural language understanding.
\newblock In \emph{Proceedings of the 33rd ACM International Conference on Information and Knowledge Management}, pages 4163--4167.

\bibitem[{Yan et~al.(2024{\natexlab{b}})Yan, Su, He, Fu, Zheng, Lyu, Wang, Wang, Wen, and Hu}]{yan2024survey}
Yibo Yan, Jiamin Su, Jianxiang He, Fangteng Fu, Xu~Zheng, Yuanhuiyi Lyu, Kun Wang, Shen Wang, Qingsong Wen, and Xuming Hu. 2024{\natexlab{b}}.
\newblock A survey of mathematical reasoning in the era of multimodal large language model: Benchmark, method \& challenges.
\newblock \emph{arXiv preprint arXiv:2412.11936}.

\bibitem[{Yan et~al.(2025{\natexlab{a}})Yan, Wang, Huo, Hu, and Wen}]{yan2025mathagent}
Yibo Yan, Shen Wang, Jiahao Huo, Xuming Hu, and Qingsong Wen. 2025{\natexlab{a}}.
\newblock Mathagent: Leveraging a mixture-of-math-agent framework for real-world multimodal mathematical error detection.
\newblock \emph{arXiv}.

\bibitem[{Yan et~al.(2024{\natexlab{c}})Yan, Wang, Huo, Li, Li, Su, Gao, Zhang, Xu, Chu et~al.}]{yan2024errorradar}
Yibo Yan, Shen Wang, Jiahao Huo, Hang Li, Boyan Li, Jiamin Su, Xiong Gao, Yi-Fan Zhang, Tianlong Xu, Zhendong Chu, and 1 others. 2024{\natexlab{c}}.
\newblock Errorradar: Benchmarking complex mathematical reasoning of multimodal large language models via error detection.
\newblock \emph{arXiv preprint arXiv:2410.04509}.

\bibitem[{Yan et~al.(2025{\natexlab{b}})Yan, Wang, Huo, Ye, Chu, Hu, Yu, Gomes, Selman, and Wen}]{yan2025position}
Yibo Yan, Shen Wang, Jiahao Huo, Jingheng Ye, Zhendong Chu, Xuming Hu, Philip~S Yu, Carla Gomes, Bart Selman, and Qingsong Wen. 2025{\natexlab{b}}.
\newblock Position: Multimodal large language models can significantly advance scientific reasoning.
\newblock \emph{arXiv preprint arXiv:2502.02871}.

\bibitem[{Yang et~al.(2025)Yang, Jimenez, Wettig, Lieret, Yao, Narasimhan, and Press}]{yang2025swe}
John Yang, Carlos Jimenez, Alexander Wettig, Kilian Lieret, Shunyu Yao, Karthik Narasimhan, and Ofir Press. 2025.
\newblock Swe-agent: Agent-computer interfaces enable automated software engineering.
\newblock \emph{Advances in Neural Information Processing Systems}, 37:50528--50652.

\bibitem[{Yang et~al.(2024{\natexlab{a}})Yang, Jimenez, Zhang, Lieret, Yang, Wu, Press, Muennighoff, Synnaeve, Narasimhan et~al.}]{yang2024swe}
John Yang, Carlos~E Jimenez, Alex~L Zhang, Kilian Lieret, Joyce Yang, Xindi Wu, Ori Press, Niklas Muennighoff, Gabriel Synnaeve, Karthik~R Narasimhan, and 1 others. 2024{\natexlab{a}}.
\newblock Swe-bench multimodal: Do ai systems generalize to visual software domains?
\newblock \emph{arXiv preprint arXiv:2410.03859}.

\bibitem[{Yang et~al.(2024{\natexlab{b}})Yang, Chu, Darwin, Han, Li, Wen, Copur-Gencturk, Tang, and Liu}]{yang2024content}
Kaiqi Yang, Yucheng Chu, Taylor Darwin, Ahreum Han, Hang Li, Hongzhi Wen, Yasemin Copur-Gencturk, Jiliang Tang, and Hui Liu. 2024{\natexlab{b}}.
\newblock Content knowledge identification with multi-agent large language models (llms).
\newblock In \emph{International Conference on Artificial Intelligence in Education}, pages 284--292. Springer.

\bibitem[{Yang et~al.(2024{\natexlab{c}})Yang, Xu, Yao, Rogers, Zhang, Intille, Shara, Gao, and Wang}]{yang2024talk2care}
Ziqi Yang, Xuhai Xu, Bingsheng Yao, Ethan Rogers, Shao Zhang, Stephen Intille, Nawar Shara, Guodong~Gordon Gao, and Dakuo Wang. 2024{\natexlab{c}}.
\newblock Talk2care: An llm-based voice assistant for communication between healthcare providers and older adults.
\newblock \emph{Proceedings of the ACM on Interactive, Mobile, Wearable and Ubiquitous Technologies}, 8(2):1--35.

\bibitem[{Yang et~al.(2024{\natexlab{d}})Yang, Liu, Gao, Xie, Li, Ouyang, Poria, Cambria, and Zhou}]{yang2024moose}
Zonglin Yang, Wanhao Liu, Ben Gao, Tong Xie, Yuqiang Li, Wanli Ouyang, Soujanya Poria, Erik Cambria, and Dongzhan Zhou. 2024{\natexlab{d}}.
\newblock Moose-chem: Large language models for rediscovering unseen chemistry scientific hypotheses.
\newblock \emph{arXiv preprint arXiv:2410.07076}.

\bibitem[{Yao et~al.(2023)Yao, Yu, Zhao, Shafran, Griffiths, Cao, and Narasimhan}]{yao2023tree}
Shunyu Yao, Dian Yu, Jeffrey Zhao, Izhak Shafran, Tom Griffiths, Yuan Cao, and Karthik Narasimhan. 2023.
\newblock Tree of thoughts: Deliberate problem solving with large language models.
\newblock \emph{Advances in neural information processing systems}, 36:11809--11822.

\bibitem[{Ye et~al.(2024{\natexlab{a}})Ye, Jiang, Wang, Li, Li, Zheng, Xie, and Huang}]{ye2024productagent}
Jingheng Ye, Yong Jiang, Xiaobin Wang, Yinghui Li, Yangning Li, Hai-Tao Zheng, Pengjun Xie, and Fei Huang. 2024{\natexlab{a}}.
\newblock Productagent: Benchmarking conversational product search agent with asking clarification questions.
\newblock \emph{arXiv preprint arXiv:2407.00942}.

\bibitem[{Ye et~al.(2023{\natexlab{a}})Ye, Li, Li, and Zheng}]{ye2023mixedit}
Jingheng Ye, Yinghui Li, Yangning Li, and Hai-Tao Zheng. 2023{\natexlab{a}}.
\newblock Mixedit: Revisiting data augmentation and beyond for grammatical error correction.
\newblock \emph{arXiv preprint arXiv:2310.11671}.

\bibitem[{Ye et~al.(2022)Ye, Li, Ma, Xie, Wu, and Zheng}]{ye2022focus}
Jingheng Ye, Yinghui Li, Shirong Ma, Rui Xie, Wei Wu, and Hai-Tao Zheng. 2022.
\newblock Focus is what you need for chinese grammatical error correction.
\newblock \emph{arXiv preprint arXiv:2210.12692}.

\bibitem[{Ye et~al.(2023{\natexlab{b}})Ye, Li, and Zheng}]{ye2023system}
Jingheng Ye, Yinghui Li, and Haitao Zheng. 2023{\natexlab{b}}.
\newblock System report for ccl23-eval task 7: Thu kelab (sz)-exploring data augmentation and denoising for chinese grammatical error correction.
\newblock In \emph{Proceedings of the 22nd Chinese National Conference on Computational Linguistics (Volume 3: Evaluations)}, pages 262--270.

\bibitem[{Ye et~al.(2023{\natexlab{c}})Ye, Li, Zhou, Li, Ma, Zheng, and Shen}]{ye2023cleme}
Jingheng Ye, Yinghui Li, Qingyu Zhou, Yangning Li, Shirong Ma, Hai-Tao Zheng, and Ying Shen. 2023{\natexlab{c}}.
\newblock Cleme: debiasing multi-reference evaluation for grammatical error correction.
\newblock \emph{arXiv preprint arXiv:2305.10819}.

\bibitem[{Ye et~al.(2024{\natexlab{b}})Ye, Qin, Li, Cheng, Qin, Zheng, Xing, Xu, Cheng, and Wei}]{ye2024excgec}
Jingheng Ye, Shang Qin, Yinghui Li, Xuxin Cheng, Libo Qin, Hai-Tao Zheng, Peng Xing, Zishan Xu, Guo Cheng, and Zhao Wei. 2024{\natexlab{b}}.
\newblock Excgec: A benchmark of edit-wise explainable chinese grammatical error correction.
\newblock \emph{arXiv preprint arXiv:2407.00924}.

\bibitem[{Ye et~al.(2025{\natexlab{a}})Ye, Qin, Li, Zheng, Wang, and Wen}]{ye2025corrections}
Jingheng Ye, Shang Qin, Yinghui Li, Hai-Tao Zheng, Shen Wang, and Qingsong Wen. 2025{\natexlab{a}}.
\newblock Corrections meet explanations: A unified framework for explainable grammatical error correction.
\newblock \emph{arXiv preprint arXiv:2502.15261}.

\bibitem[{Ye et~al.(2025{\natexlab{b}})Ye, Wang, Zou, Yan, Wang, Zheng, Xu, King, Yu, and Wen}]{ye2025position}
Jingheng Ye, Shen Wang, Deqing Zou, Yibo Yan, Kun Wang, Hai-Tao Zheng, Zenglin Xu, Irwin King, Philip~S Yu, and Qingsong Wen. 2025{\natexlab{b}}.
\newblock Position: Llms can be good tutors in foreign language education.
\newblock \emph{arXiv preprint arXiv:2502.05467}.

\bibitem[{Ye et~al.(2024{\natexlab{c}})Ye, Xu, Li, Cheng, Song, Zhou, Zheng, Shen, and Su}]{ye2024cleme2}
Jingheng Ye, Zishan Xu, Yinghui Li, Xuxin Cheng, Linlin Song, Qingyu Zhou, Hai-Tao Zheng, Ying Shen, and Xin Su. 2024{\natexlab{c}}.
\newblock Cleme2. 0: Towards more interpretable evaluation by disentangling edits for grammatical error correction.
\newblock \emph{arXiv preprint arXiv:2407.00934}.

\bibitem[{Yu et~al.(2024)Yu, Baker, Chen, Herb, Gou, Adu-Ampratwum, Ning, and Sun}]{yu2024tooling}
Botao Yu, Frazier~N Baker, Ziru Chen, Garrett Herb, Boyu Gou, Daniel Adu-Ampratwum, Xia Ning, and Huan Sun. 2024.
\newblock Tooling or not tooling? the impact of tools on language agents for chemistry problem solving.
\newblock \emph{arXiv preprint arXiv:2411.07228}.

\bibitem[{Yuan et~al.(2024)Yuan, Cao, Jiang, Kang, Lin, Song, Yan, Sun, Liu et~al.}]{yuan2024can}
Weikang Yuan, Junjie Cao, Zhuoren Jiang, Yangyang Kang, Jun Lin, Kaisong Song, Pengwei Yan, Changlong Sun, Xiaozhong Liu, and 1 others. 2024.
\newblock Can large language models grasp legal theories? enhance legal reasoning with insights from multi-agent collaboration.
\newblock \emph{arXiv preprint arXiv:2410.02507}.

\bibitem[{Yue et~al.(2024)Yue, Lyu, Mifdal, Suh, Zhang, and Yao}]{yue2024mathvc}
Murong Yue, Wenhan Lyu, Wijdane Mifdal, Jennifer Suh, Yixuan Zhang, and Ziyu Yao. 2024.
\newblock Mathvc: An llm-simulated multi-character virtual classroom for mathematics education.
\newblock \emph{arXiv preprint arXiv:2404.06711}.

\bibitem[{Zaiane(2002)}]{zaiane2002building}
Osmar~R Zaiane. 2002.
\newblock Building a recommender agent for e-learning systems.
\newblock In \emph{International Conference on Computers in Education, 2002. Proceedings.}, pages 55--59. IEEE.

\bibitem[{Zha et~al.(2024)Zha, Qiao, Hu, Li, Gong, and Xu}]{zha2024designing}
Siyu Zha, Yuehan Qiao, Qingyu Hu, Zhongsheng Li, Jiangtao Gong, and Yingqing Xu. 2024.
\newblock Designing child-centric ai learning environments: Insights from llm-enhanced creative project-based learning.
\newblock \emph{arXiv preprint arXiv:2403.16159}.

\bibitem[{Zhai et~al.(2021)Zhai, Chu, Chai, Jong, Istenic, Spector, Liu, Yuan, and Li}]{zhai2021review}
Xuesong Zhai, Xiaoyan Chu, Ching~Sing Chai, Morris Siu~Yung Jong, Andreja Istenic, Michael Spector, Jia-Bao Liu, Jing Yuan, and Yan Li. 2021.
\newblock A review of artificial intelligence (ai) in education from 2010 to 2020.
\newblock \emph{Complexity}, 2021(1):8812542.

\bibitem[{Zhang et~al.(2024{\natexlab{a}})Zhang, Li, Wang, Zhang, Zhou, and Qiu}]{zhang2024speechagents}
Dong Zhang, Zhaowei Li, Pengyu Wang, Xin Zhang, Yaqian Zhou, and Xipeng Qiu. 2024{\natexlab{a}}.
\newblock Speechagents: Human-communication simulation with multi-modal multi-agent systems.
\newblock \emph{arXiv preprint arXiv:2401.03945}.

\bibitem[{Zhang et~al.(2024{\natexlab{b}})Zhang, Li, Li, Shi, and Jin}]{zhang2024codeagent}
Kechi Zhang, Jia Li, Ge~Li, Xianjie Shi, and Zhi Jin. 2024{\natexlab{b}}.
\newblock Codeagent: Enhancing code generation with tool-integrated agent systems for real-world repo-level coding challenges.
\newblock \emph{arXiv preprint arXiv:2401.07339}.

\bibitem[{Zhang et~al.(2025{\natexlab{a}})Zhang, Dilling, Gondelman, Lyngdorf, Lindsay, and Bjerva}]{zhang2025sefl}
Mike Zhang, Amalie~Pernille Dilling, L{\'e}on Gondelman, Niels Erik~Ruan Lyngdorf, Euan~D Lindsay, and Johannes Bjerva. 2025{\natexlab{a}}.
\newblock Sefl: Harnessing large language model agents to improve educational feedback systems.
\newblock \emph{arXiv preprint arXiv:2502.12927}.

\bibitem[{Zhang et~al.(2025{\natexlab{b}})Zhang, Tang, Zang, Pei, Liang, Zhao, and Zhao}]{zhang2025let}
Rongsheng Zhang, Jiji Tang, Chuanqi Zang, Mingtao Pei, Wei Liang, Zeng Zhao, and Zhou Zhao. 2025{\natexlab{b}}.
\newblock Let storytelling tell vivid stories: A multi-modal-agent-based unified storytelling framework.
\newblock \emph{Neurocomputing}, 622:129316.

\bibitem[{Zhang et~al.(2025{\natexlab{c}})Zhang, Zhang, Sun, Xiao, Yang, and Luo}]{zhang2025eduplanner}
Xueqiao Zhang, Chao Zhang, Jianwen Sun, Jun Xiao, Yi~Yang, and Yawei Luo. 2025{\natexlab{c}}.
\newblock Eduplanner: Llm-based multi-agent systems for customized and intelligent instructional design.
\newblock \emph{IEEE Transactions on Learning Technologies}.

\bibitem[{Zhang et~al.(2025{\natexlab{d}})Zhang, Li, Song, Sun, Xu, and Wen}]{zhang2025correctness}
Yi-Fan Zhang, Hang Li, Dingjie Song, Lichao Sun, Tianlong Xu, and Qingsong Wen. 2025{\natexlab{d}}.
\newblock From correctness to comprehension: Ai agents for personalized error diagnosis in education.
\newblock \emph{arXiv preprint arXiv:2502.13789}.

\bibitem[{Zhang et~al.(2023)Zhang, Li, Cui, Cai, Liu, Fu, Huang, Zhao, Zhang, Chen et~al.}]{zhang2023siren}
Yue Zhang, Yafu Li, Leyang Cui, Deng Cai, Lemao Liu, Tingchen Fu, Xinting Huang, Enbo Zhao, Yu~Zhang, Yulong Chen, and 1 others. 2023.
\newblock Siren’s song in the ai ocean: A survey on hallucination in large language models.
\newblock \emph{arXiv preprint arXiv:2309.01219}, 2(5).

\bibitem[{Zhang et~al.(2024{\natexlab{c}})Zhang, Zhang, Li, Gao, Wang, Lu, Zhao, Qiao, and Shao}]{zhang2024psysafe}
Zaibin Zhang, Yongting Zhang, Lijun Li, Hongzhi Gao, Lijun Wang, Huchuan Lu, Feng Zhao, Yu~Qiao, and Jing Shao. 2024{\natexlab{c}}.
\newblock Psysafe: A comprehensive framework for psychological-based attack, defense, and evaluation of multi-agent system safety.
\newblock \emph{arXiv preprint arXiv:2401.11880}.

\bibitem[{Zhang et~al.(2024{\natexlab{d}})Zhang, Bo, Ma, Li, Chen, Dai, Zhu, Dong, and Wen}]{zhang2024survey}
Zeyu Zhang, Xiaohe Bo, Chen Ma, Rui Li, Xu~Chen, Quanyu Dai, Jieming Zhu, Zhenhua Dong, and Ji-Rong Wen. 2024{\natexlab{d}}.
\newblock A survey on the memory mechanism of large language model based agents.
\newblock \emph{arXiv preprint arXiv:2404.13501}.

\bibitem[{Zhao et~al.(2025)Zhao, Ma, Xu, Kong, and Deng}]{zhao2025biomaze}
Haiteng Zhao, Chang Ma, FangZhi Xu, Lingpeng Kong, and Zhi-Hong Deng. 2025.
\newblock Biomaze: Benchmarking and enhancing large language models for biological pathway reasoning.
\newblock \emph{arXiv preprint arXiv:2502.16660}.

\bibitem[{Zheng et~al.(2024)Zheng, Hong, Liu, Wang, Su, Liang, and Wu}]{zheng2024fine}
Jiawei Zheng, Hanghai Hong, Feiyan Liu, Xiaoli Wang, Jingsong Su, Yonggui Liang, and Shikai Wu. 2024.
\newblock Fine-tuning large language models for domain-specific machine translation.
\newblock \emph{arXiv preprint arXiv:2402.15061}.

\bibitem[{Zheng et~al.(2025{\natexlab{a}})Zheng, Jiang, Gu, Li, Wang, and Zhang}]{zheng2025teaching}
Longwei Zheng, Fei Jiang, Xiaoqing Gu, Yuanyuan Li, Gong Wang, and Haomin Zhang. 2025{\natexlab{a}}.
\newblock Teaching via llm-enhanced simulations: Authenticity and barriers to suspension of disbelief.
\newblock \emph{The Internet and Higher Education}, 65:100990.

\bibitem[{Zheng et~al.(2025{\natexlab{b}})Zheng, Liao, Fu, Lei, Lyu, Jiang, Ren, Chen, Wang, Li et~al.}]{zheng2025mllms}
Xu~Zheng, Chenfei Liao, Yuqian Fu, Kaiyu Lei, Yuanhuiyi Lyu, Lutao Jiang, Bin Ren, Jialei Chen, Jiawen Wang, Chengxin Li, and 1 others. 2025{\natexlab{b}}.
\newblock Mllms are deeply affected by modality bias.
\newblock \emph{arXiv preprint arXiv:2505.18657}.

\bibitem[{Zhong et~al.(2023)Zhong, Guo, Gao, and Wang}]{zhong2023memorybank}
Wanjun Zhong, Lianghong Guo, Qiqi Gao, and Yanlin Wang. 2023.
\newblock Memorybank: Enhancing large language models with long-term memory.
\newblock \emph{arXiv preprint arXiv:2305.10250}.

\bibitem[{Zhou et~al.(2023)Zhou, Xu, Zhu, Zhou, Lo, Sridhar, Cheng, Ou, Bisk, Fried et~al.}]{zhou2023webarena}
Shuyan Zhou, Frank~F Xu, Hao Zhu, Xuhui Zhou, Robert Lo, Abishek Sridhar, Xianyi Cheng, Tianyue Ou, Yonatan Bisk, Daniel Fried, and 1 others. 2023.
\newblock Webarena: A realistic web environment for building autonomous agents.
\newblock \emph{arXiv preprint arXiv:2307.13854}.

\bibitem[{Zhu et~al.(2024)Zhu, Wang, Gao, Xu, Wang, Liu, Wang, Jin, Pang, Wen et~al.}]{zhu2024recommender}
Xi~Zhu, Yu~Wang, Hang Gao, Wujiang Xu, Chen Wang, Zhiwei Liu, Kun Wang, Mingyu Jin, Linsey Pang, Qingsong Wen, and 1 others. 2024.
\newblock Recommender systems meet large language model agents: A survey.
\newblock \emph{Available at SSRN 5062105}.

\bibitem[{Zhuang et~al.(2023)Zhuang, Yu, Wang, Sun, and Zhang}]{zhuang2023toolqa}
Yuchen Zhuang, Yue Yu, Kuan Wang, Haotian Sun, and Chao Zhang. 2023.
\newblock Toolqa: A dataset for llm question answering with external tools.
\newblock \emph{Advances in Neural Information Processing Systems}, 36:50117--50143.

\bibitem[{Zou et~al.(2025)Zou, Ye, Liu, Wu, Xu, Li, Zheng, An, Wei, and Xu}]{zou2025revisiting}
Deqing Zou, Jingheng Ye, Yulu Liu, Yu~Wu, Zishan Xu, Yinghui Li, Hai-Tao Zheng, Bingxu An, Zhao Wei, and Yong Xu. 2025.
\newblock Revisiting classification taxonomy for grammatical errors.
\newblock \emph{arXiv preprint arXiv:2502.11890}.

\end{thebibliography}
